# Causally-informed Deep Learning to Improve Climate Models and Projections

Fernando Iglesias-Suarez[1], Pierre Gentine[2,3], Breixo Solino-Fernandez[1], Tom Beucler[4], Michael Pritchard[5,6], Jakob Runge[7,8], and Veronika Eyring[1,9]

[1]Deutsches Zentrum für Luft- und Raumfahrt e.V. (DLR), Institute of Atmospheric Physics, Oberpfaffenhofen, Germany
[2]Department of Earth and Environmental Engineering, Center for Learning the Earth with Artificial intelligence and Physics (LEAP), Columbia University, New York, USA
[3]Earth and Environmental Engineering, Earth and Environmental Sciences, Learning the Earth with Artificial intelligence and Physics (LEAP) Science and Technology Center, Columbia University, New York, USA
[4]University of Lausanne, Institute of Earth Surface Dynamics, Lausanne, Switzerland
[5]University of California, Department of Earth System Science, Irvine, USA
[6]NVIDIA Corporation, Santa Clara, USA
[7]Deutsches Zentrum für Luft- und Raumfahrt e.V. (DLR), Institute of Data Science, Jena, Germany
[8]Technische Universität Berlin, Institute of Computer Engineering and Microelectronics, Berlin, Germany
[9]University of Bremen, Institute of Environmental Physics (IUP), Bremen, Germany

**Key Points:**

- Causal discovery unveils causal drivers of subgrid-scale processes across climates
- The causally-informed hybrid model runs stably and generates a climate close to the original high-resolution simulation
- Spurious correlations are evident in the non-causal parameterization, leading to underestimate feature importance of physical drivers

Corresponding author: F. Iglesias-Suarez, `figlesua@gmail.com`

**Abstract**
Climate models are essential to understand and project climate change, yet long-standing biases and uncertainties in their projections remain. This is largely associated with the representation of subgrid-scale processes, particularly clouds and convection. Deep learning can learn these subgrid-scale processes from computationally expensive storm-resolving models while retaining many features at a fraction of computational cost. Yet, climate simulations with embedded neural network parameterizations are still challenging and highly depend on the deep learning solution. This is likely associated with spurious non-physical correlations learned by the neural networks due to the complexity of the physical dynamical system. Here, we show that the combination of causality with deep learning helps removing spurious correlations and optimizing the neural network algorithm. To resolve this, we apply a causal discovery method to unveil causal drivers in the set of input predictors of atmospheric subgrid-scale processes of a superparameterized climate model in which deep convection is explicitly resolved. The resulting causally-informed neural networks are coupled to the climate model, hence, replacing the superparameterization and radiation scheme. We show that the climate simulations with causally-informed neural network parameterizations retain many convection-related properties and accurately generate the climate of the original high-resolution climate model, while retaining similar generalization capabilities to unseen climates compared to the non-causal approach. The combination of causal discovery and deep learning is a new and promising approach that leads to stable and more trustworthy climate simulations and paves the way towards more physically-based causal deep learning approaches also in other scientific disciplines.

**Plain Language Summary**

Climate models have biases compared to observations that have been present for a long time because certain processes, like convection, are only approximated using simplified methods. Neural networks can better represent these processes, but often learn incorrect connections leading to unreliable results and climate model crashes. To solve this, we used a new method that informs neural networks with causal drivers, therefore, respecting the underlying physical processes. By doing so, we developed more reliable and trustworthy neural networks, allowing us to accurately represent the climate of the original high-resolution simulation on which these neural networks were trained.

# 1 Introduction

Our understanding of the climate system and how it may change in the future under different scenarios has greatly improved thanks to climate models (IPCC, 2021b). Yet, systematic biases still plague current climate models compared to observations (Flato et al., 2013; Eyring, Gillett, et al., 2021) and limit their ability to accurately project climate change at global and regional scales (Tebaldi et al., 2021; Lee et al., 2021). Many important processes determining the Earth's climate occur at scales smaller than current climate models grid size, typically ranging 50–100 kilometers horizontally (IPCC, 2021a). The effect of these subgrid-scale or unresolved processes, such as clouds and convection, on the system are approximated via physical parameterizations in current models, and are a key source of the uncertainty in climate projections (Schneider et al., 2017; Gentine et al., 2021).

High-resolution storm-resolving models (SRMs), run at a horizontal scale of few kilometers, explicitly represent deep convection and dynamics of convective storms, and alleviate a number of biases present in coarser climate models (Sherwood et al., 2014; Bock et al., 2020). For instance, they show improvements in representing the Intertropical Convergence Zone (ITCZ) (Klocke et al., 2017), storm-tracks and precipitation (Stevens et al., 2019), as well as subseasonal variability (Rasp et al., 2018). Accurate represen-tation of convective processes is also essential to capture Earth's system feedbacks in a changing climate, like cloud-radiation feedbacks (Bony et al., 2015). Yet, global SRMs simulations are only possible over short periods of time (months) due to their staggering computational costs, whereas climate research requires a number of realizations over hundred of years (Schneider et al., 2017).

Machine learning (ML) approaches, and in particular deep learning (DL) methods, have shown great potential in learning explicitly resolved small-scale processes such as deep convection from SRM simulations (Eyring, Mishra, et al., 2021; Gentine et al., 2018, 2021; Grundner et al., 2022) and represent them in coarser resolution models. Hybrid models, i.e., ML-based parameterizations coupled to a host climate model, have shown great performance simulating the climate of the original SRM in terms of mean state and its variability (e.g., tropical waves) (Rasp et al., 2018; Yuval & O'Gorman, 2020; Watt-Meyer et al., 2021; Bretherton et al., 2022; Wang et al., 2022). ML methods for Earth system modeling is an active area of research. Particularly challenging to address are the poor representation of unseen climates and regimes (i.e., generalization capabilities) (O'Gorman & Dwyer, 2018; Scher & Messori, 2019; Grundner et al., 2022), and hybrid modeling instabilities associated with the interaction between the ML-based algorithm and the dynamical core of the host climate model (Brenowitz, Beucler, et al., 2020). Training the ML algorithm directly online, coupled to the climate model, (Lopez-Gomez et al., 2022; Frezat et al., 2022), crude ablation (Brenowitz & Bretherton, 2019) and deeper DL algorithms (Rasp et al., 2018) have been proposed to overcome hybrid ML modeling instabilities. Yet, the causes of such instabilities are not fully understood. Our working hypothesis is that ML-based parameterizations can accurately reproduce subgrid-scale processes using non-causal relationships (i.e., mere correlations), and these correlations might be overfitting the training dataset (Brenowitz, Henn, et al., 2020). In a nutshell, ML algorithms can skillfully learn a given task for the wrong reasons using spurious non-physical relationships, but may struggle in conditions deviating from the training data in which causes and effects might differ from the initial correlations present in the training data (Brenowitz, Henn, et al., 2020).

Integrating domain knowledge in the form of causal relationships (i.e., inductive bias) is a recent and emerging theme in machine learning research (Pearl & Mackenzie, 2018; Runge, Bathiany, et al., 2019; Schölkopf et al., 2021) to overcome shortcomings of standard ML methods, which predictions are mostly based on correlations between predictors and predictands. While correlation is a statistical relationship between two variables (i.e., where a change in one variable is associated with a change in the other variable), causality is the relationship between an action (the cause) and its outcome (the effect). As an example, for data coming from a causal model $X^1 \leftarrow X^2 \rightarrow Y$, an ML algorithm may learn to predict $Y$ from both $X^1$ and $X^2$. However, such a prediction would fail if the ML method is employed under changing environments where the confounder $X^1$ is in a different state. This can easily be the case when an ML algorithm learns an atmospheric physical process. Owing to the strong correlation induced by convective processes in the atmospheric profile environment, surface precipitation, for example, can be influenced not only by conditions in the troposphere but also by upper tropospheric moisture (confounder) (Brenowitz & Bretherton, 2019). Causal discovery methods aim to discover such causal relationships from data (Runge, Bathiany, et al., 2019; Runge, Nowack, et al., 2019), going beyond simple correlations. The goal of our study is to merge the power of causal discovery with the data exploitation capacity of neural networks, and investigate whether such a causally-informed neural network can help better understand and predict (subgrid) physical mechanisms in the atmosphere.

We build on existing data-driven subgrid parameterization work (Rasp et al., 2018) and combine causal discovery and deep learning, using the same high-resolution modeling data, to improve DL-based parameterizations. The task of the causal discovery algorithm is to unveil the causal drivers of the subgrid-scale processes respecting the un-derlying physical mechanisms. The causally-informed neural network algorithms have two steps. We first identify causal drivers of subgrid-scale processes using a causal discovery method based on conditional independence tests. Then we build novel causally-informed neural network algorithms, in which subgrid-scale processes are learned from the causal drivers. In other words, we build sparser (lower dimensional) neural networks in which non-causal connections are dropped. We demonstrate several key aspects of this novel method: 1) causal discovery removes spurious links; 2) causal drivers are "climate invariant" (i.e., robust across colder and warmer climates (Beucler et al., 2021)); and 3) the causally-informed hybrid model accurately represents the climate of the original high-resolution climate model, retaining many convection-related properties. We finish by discussing a potential broader role of causal discovery in the context of machine learning for physical sciences, and key remaining challenges for future work.

**Table 1.** Summary of the Model Simulations.

| Climate models | Parameterizations | Causal-threshold |
| --- | --- | --- |
| SPCAM | SP component (2–D SRM) | — |
| Non-causalNNCAM | Non-causalNN | — |
| Causal$_{0.59}$NNCAM | Causally-informed$_{0.59}$NN | quantile 0.59[a] |
| CausalNNCAM | Causally-informedNN | quantile optimized[b] |

[a]Single optimized causal-threshold for all outputs.
[b]Varying optimized causal-threshold (see Supplementary Information).

## 2 Causally-Informed Hybrid Modeling

We extend previous work (Gentine et al., 2018; Rasp et al., 2018) to build a causally-informed hybrid climate model (see Table 1). Figure 1 shows a schematic overview of the causally-informed neural network approach. We use an aquaplanet (i.e., ocean only without topography) setup of the Superparameterized Community Atmosphere Model v3.0 (SPCAM) (Collins et al., 2006). The model extends from the surface to the upper stratosphere (3.5 hPa) with 30 vertical levels and includes a horizontal resolution of 2.8°×2.8° (latitude by longitude). Stationary –no seasonality but with diurnal cycle– zonal mean sea surface temperatures are imposed using a realistic equator-to-pole gradient (Andersen & Kuang, 2012). The time step of the climate model component (CAM) is 30 minutes. The superparameterization component (SP) is a 2–D SRM embedded in each grid column, explicitly resolving most of deep convection but parameterizing turbulence and cloud microphysics (M. F. Khairoutdinov & Randall, 2001; Pritchard et al., 2014; Pritchard & Bretherton, 2014). For consistency with the former study (Rasp et al., 2018), the SP component uses eight 4–km-wide meridionally oriented columns (west to east), and time steps of 20 seconds. SPCAM alleviates a number of climate model biases (Oueslati & Bellon, 2015), including a more realistic Madden-Julian oscillation and a single ITCZ, as well as better representation of precipitation extremes (Benedict & Randall, 2009; Arnold & Randall, 2015; Kooperman et al., 2016b, 2016a, 2018).

The task of the neural networks (NNs) is to learn subgrid-scale processes (output predictands) as represented by the SP component given the environmental conditions (input predictors) of the climate model, CAM. The training data are column-based values of the model's subgrid physics package, which includes the SP subgrid resolution of convective transport, turbulence and radiation scheme, with a few omissions (condensed water species and ozone). The inputs ($X$) are column-wise values of temperature $T(p)\,[K]$, specific humidity $q(p)\,[kg\,kg^{-1}]$, and meridional wind $V(p)\,[m\,s^{-1}]$ at each column level, surface pressure $P_{srf}\,[Pa]$, incoming solar radiation $Q_{sol}$ at the top of the atmosphere, as well as sensible- and latent-heat fluxes at the surface, $Q_{sen}$ and $Q_{lat}$ in $[W\,m^{-2}]$, respectively. The outputs ($Y$) comprise: heating tendencies $\Delta T_{phy}(p)$ (including convection and radiative heating rates) $[K\,s^{-1}]$; and moistening tendencies $\Delta q_{phy}(p)\,[kg\,kg^{-1}s^{-1}]$ at each model level; net shortwave and longwave radiative heat fluxes at the model top and at the surface ($Q_{sw}^{top}$, $Q_{lw}^{top}$, $Q_{sw}^{srf}$ and $Q_{lw}^{srf}$ respectively) $[W\,m^{-2}]$, and surface precipitation $P\,[kg\,m^{-2}\,d^{-1}]$. Only the heating and moistening tendencies are coupled to the climate model's dynamical core, with the other outputs being diagnostics. Two data-driven DL parameterizations are considered: non-causal; and causally-informed (see Parameterizations column in Table 1). In both cases, and based on previous work (Rasp et al., 2018), we use fully connected feedforward NNs, with 9 hidden layers and 256 units per layer (around 0.5 million parameters). The NNs are optimized by minimizing the loss defined as the mean squared error of the prediction compared to the SPCAM "truth" value. We use 3 months of an SPCAM simulation for training, validation and test sets (each set approximately including 45 million samples using every model time step and grid column). Note the training set is shuffled in time and space (grid columns). See Supplementary Information for additional details.

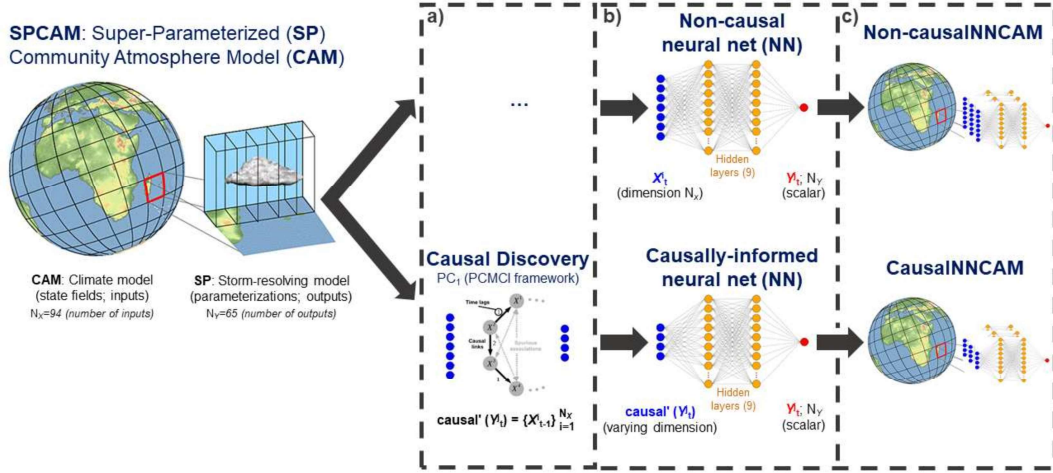

**Figure 1.** Schematic overview of the causally-informed neural network approach. The Super-parameterized Community Atmosphere Model (SPCAM) is used to learn subgrid-scale processes ($Y$) as represented by the SP component given the environmental conditions ($X$) of the climate model, CAM. Two data-driven parameterizations are considered: (top) non-causal; and (bottom) causally-informed. For the causally-informed approach, **a**) we use a causal discovery algorithm, the $PC_1$ phase of the PCMCI method (Runge, Nowack, et al., 2019), to prune the fully connected input vector and eliminate non-physical spurious inputs and connections. Therefore, **b**) we develop 65 separate single-output NNs, each one having a specific subset of causal inputs, $causal'(Y_t^j)$, with a varying input vector length. The main difference between the analogous Causally-informedNN and Non-causalNN parameterizations is that the latter always includes an input vector of length 94 (i.e., full set of environmental conditions). Finally, **c**) we couple both the Non-causalNN parameterization and the new Causally-informedNN parameterization of SPCAM physics within CAM, resulting in the Non-causalNNCAM and CausalNNCAM models respectively (Table 1). The causal discovery diagram was adapted from Runge et al. (Runge, Nowack, et al., 2019) and the sketch of SPCAM from Prein et al. (Prein et al., 2015).

We develop 65 separate single-output NNs for both cases, causal and non-causal parameterizations (corresponding to 30 vertical levels for heating and moistening rates, plus net shortwave and longwave radiative heat fluxes at the model top and at the surface, and surface precipitation), rather than a single NN for the entire column (Han et al., 2020). In this way, each causally-informed NN has a specific subset of causal inputs (drivers) obtained during the causal discovery phase, and therefore a varying input vector length (see below). The main difference between the analogous causally-informed NNs and non-causal NNs is that the latter always include an input vector of length 94 (i.e., 3 times the number of levels for temperature, humidity, and wind, as well as 2–D fields of surface pressure, sensible and latent heat fluxes, and top of the atmosphere incoming solar radiation) (see Supplementary Information). For the causally-informed NNs, our goal is to prune the fully connected input vector to eliminate non-physical spurious inputs and connections. From a causal perspective, the setup here is simplified because the inputs and outputs are known to be separated in time based on the SP climate model's structure. Hence, we can utilize a causal discovery selection algorithm that removes those inputs that are conditionally independent of the output, thus, providing no additional information to predict the output.

While no causal discovery method is infallible, we can gain a deeper understanding of how a dynamical system works and develop more informed algorithms based on causal evidence rather than simply relying on correlations or associations between variables (i.e., environmental conditions driving subgrid-scale processes). Our selection algorithm is the PC$_1$ phase of the PCMCI method (Runge, Nowack, et al., 2019), which is based on the PC algorithm (Spirtes & Glymour, 1991), and the Momentary Conditional Independence (MCI) test. PCMCI, and its different flavors, have been widely used in recent years in climate sciences, such as for better understanding teleconnections in the Earth system (Runge et al., 2014; Kretschmer et al., 2016, 2018; Siew et al., 2020) and their pathways (Runge et al., 2015; Kretschmer et al., 2021; Karmouche et al., 2023; Galytska et al., 2023) or to investigate land-atmosphere interactions (Krich et al., 2020).

PC$_1$ starts by a fully connected matrix between all inputs and outputs. This initializes the causal drivers to $causal_g(Y_t^j) = \{X_{t-1}^i\}_{i=1}^{N_X}$, where $N_X$ is the number of all potential drivers across the different vertical levels and g refers to each column of the model grid. Then PC$_1$ tests whether each input ($X_{t-1}^i$) is conditionally independent of an output ($Y_t^j$) given selected subsets of the other correlated inputs in the dataset ($causal_g(Y_t^j)$). If two variables are found to be conditionally independent, it is inferred that there is no direct causal relationship between them. Specifically, it removes drivers $X_{t-1}^i$ from $causal_g(Y_t^j)$ if they are conditionally independent (irrelevant or redundant) of $Y_t^j$ given subsets $S_k \subset causal_g(Y_t^j)$ whose cardinality $k$ iteratively increases. For $k = 0$, all $X_{t-1}^i$ with $X_{t-1}^i \perp$ $\perp Y_t^j$, where $\perp\!\!\!\perp$ refers to conditional independence, are removed. For $k = 1$, those with $X_{t-1}^i \perp\!\!\!\perp Y_t^j | S_1$ are removed, where $S_1$ is the strongest driver (as measured by their absolute partial correlation value) from the previous step. For $k = 2$, those with $X_{t-1}^i \perp\!\!\!\perp Y_t^j | S_2$ are removed, where $S_2$ are the two strongest drivers (regarding the individual absolute partial correlation values), among the remaining drivers, excluding $X_{t-1}^i$, from the previous step. A simple forward-selection method would always keep the strongest driver at each iteration step, while our approach re-tests them conditional on the remaining $k$ strongest drivers. This procedure continues until the algorithm converges and stops when there are no more possible combinations $S_k$, that is, if the cardinality of $S_k$ is equal to the number of remaining drivers. We note that the iterative process of repeatedly testing each potential causal driver of an output against efficiently chosen subsets of the other potential inputs is what establishes $PC_1$ as causal under the assumptions discussed below. Here conditional independence is tested using partial correlation because we may reasonably assume that at a short 30 min time-scale, even non-linear relations are sufficiently well captured by a linear model. The independence test is based on a standard significance level $\alpha_{pc} = 0.01$.

Our goal here is primarily to remove spurious inputs. A causal interpretation of $causal_g(Y_t^j)$ using the above algorithm rests on the following assumptions: causal sufficiency (all common causes are observed), the Markov condition (dependence must be due to causal connectedness), faithfulness (independencies are not by coincidence but structural, therefore, follow the Markov condition). This approach yields a set of causal drivers $causal_g(Y_t^j)$ for each variable $Y^j$ at every column $g$ and every vertical level. Because we want these drivers to generalize across all columns (across space and time), we define "robust" causal drivers as $causal'(Y_t^j) = \{X_{t-1}^i : P(X_{t-1}^i \rightarrow Y_t^j \in causal_g(Y_t^j)) > q\}$. Specifically, we only consider causal drivers $X_{t-1}^i$ for a given output $Y_t^j$, those whose probability $P$ of being causally-linked to the output across the 8,192 latitude and longitude columns is at least quantile $q$. This is also the main idea behind the invariant causal prediction approach (Peters et al., 2016). We explore two cases to choose $q$: a single optimized quantile-threshold of $q = 0.59$ for all outputs; and a varying quantile-threshold optimized for each output separately (quantile-threshold optimization is based on causally-informed NNs offline performance; see Supplementary Information). $q$ is the primary hyperparameter of the algorithm. A too loose threshold would lead to keeping non-physical spurious inputs similar to a fully connected (non-causal) feedforward NN. A too strict threshold would lead to neglecting some important input features (key drivers) and thus having poor predictive skill. The results presented here are based on the varying quantile-threshold (see Supplementary Information for the single optimized quantile-threshold results).

Finally, to test the value of our causal discovery approach ($PC_1$) in removing potential spurious inputs-to-outputs links, we explore two additional feature selection approaches. First, we compare $PC_1$ with a baseline (non-causal) correlation method, which follows the column-wise approach. Second, linear Lasso regression is applied to the entire training set without separating each location (Tibshirani, 1996), which under the above causal assumptions may also work as a complementary method for selecting causal features. Although we choose in this work the $PC_1$ algorithm and explore linear Lasso regression, we note there are a number of causal discovery methods that may well be suitable (Runge et al., 2023).

## 3 Results

### 3.1 Causal Discovery

Based on the causal discovery algorithm, we can investigate the main drivers of the subgrid-scale processes as represented by the SPCAM model. Causal drivers of SPCAM's parameterizations inferred by our causal discovery algorithm are in agreement with cur-rent physics understanding (Fig. 2; Fig. S1). Causal matrices, similar to transilient matrices (Stull, 1993; Romps & Kuang, 2011), reveal largest coefficients –ratio of the causal drivers appearance across the model's grid– on the diagonal, meaning that key direct drivers are primarily local in the vertical. This is especially strong for humidity, which is known to be a key regulator of convective mixing and updraft buoyancy through lateral entrainment (Stommel, 1951; Warner, 1970; de Rooy et al., 2013). Nonetheless, the causal matrices reveal the influence of the lower troposphere ($p > 600$ hPa) on the profile. This is expected since convective processes and buoyant plumes originate from the boundary layer, affecting the troposphere above. In particular, boundary-layer and lower troposphere temperature is causally-linked (driver) to moistening and heating rates throughout the troposphere. Several studies have shown that cold pools induced by unsaturated downdrafts organize the boundary layer (Del Genio & Wu, 2010; Kuang & Bretherton, 2006; M. Khairoutdinov & Randall, 2006; Mapes & Neale, 2011), and organizing convection leads to changes in atmospheric heating and moisteing tendencies and precipitation (Tompkins, 2001; Muller & Bony, 2015). Furthermore, heating rates in the upper-troposphere and lower-stratosphere (∼100–300 hPa) are associated with mid-tropospheric moistening, where deep convection can have a substantial radiative effect due to cirrus clouds. Incoming solar radiation is the most important driver of heating rates because of its regulation of the diurnal cycle. Interestingly, precipitation at the surface is strongly causally-linked by environmental conditions from the lower to the middle troposphere, and therefore associated with convective processes and rain reevaporation, consistent with the strong relationship between precipitation and the bulk temperature and moisture in the lower troposphere (Del Genio, 2012; D'Andrea et al., 2014; Ahmed & Neelin, 2018).

Our causal feature selection methodologies outperform a more naïve baseline feature selection approach removing potential spurious inputs-to-outputs links. Using a simple correlation method, we find that correlations across the atmospheric profile are largely non-local and redundant among state fields (i.e., less physical; see Fig. S2). This is due to the large inter-correlation in the atmospheric profile associated with convective processes. These strong correlations across levels would nonetheless include potential spurious links and primarily define the strength of the neural network connections. Moreover, using simple correlations to optimize the connectivity matrices is challenging, since either a number of outputs lack input links (e.g., upper tropospheric moisture; Fig. S2), or the system is quasi-fully connected (not shown). Furthermore, we use linear Lasso regression (Tibshirani, 1996), which under the above causal assumptions (sufficiency, Markov, and faithfulness) may work as a causal feature selection method. While selected features show largest Lasso coefficients on the diagonal (Fig. S3), meaning it captures that key direct physical drivers are primarily local in the vertical, there are clear spurious features (e.g., moistening and heating tendencies in the lower troposphere are associated with environmental conditions in the stratosphere; see also Section 3.3 and Fig. S9). While the correlationally-informed parameterization performs sub-optimally compared to the reference non-causal case (particularly in the lower troposphere), lasso-informed parameterization shows in general similar skills (Fig. S7). Causal discovery is arguably a more complex method that rests on some expert knowledge of the physical problem (allowing us to choose a suitable causal algorithm and its setup), and on a number of mathematical assumptions. However, it goes beyond standard feature selection approaches and helps further remove spurious links, as demonstrated by $PC_1$ and linear Lasso regression.

As a further test of the credibility of our causal feature selection methodology and its stability with a changing input distribution, we also explore its sensitivity to climate change (Galytska et al., 2023; Karmouche et al., 2023). Thermodynamic features driving different atmospheric processes are "climate invariant", i.e., they govern the same processes regardless of the climate state of the system, as physics does not change with climate change. For example, whatever the state of the climate system, we expect the key direct drivers of heating and moistening tendencies to remain local, though deep con-vection affects them non-locally throughout the troposphere. Reassuringly, we find that causal drivers of subgrid-scale processes as inferred by our causal discovery algorithm in SPCAM are consistent across climates for both, global 4 K sea surface temperatures cooling and warming (only around 5 % inconsistent non-local causal drivers in the vertical, Fig. 3). This result suggests that causal discovery helps unveil the most direct key drivers of smaller-scale processes represented by SPCAM and remove some of the confounding effects present when using neural networks, that would otherwise affect their performance due to spurious inter-correlations across the vertical profile (shown below).

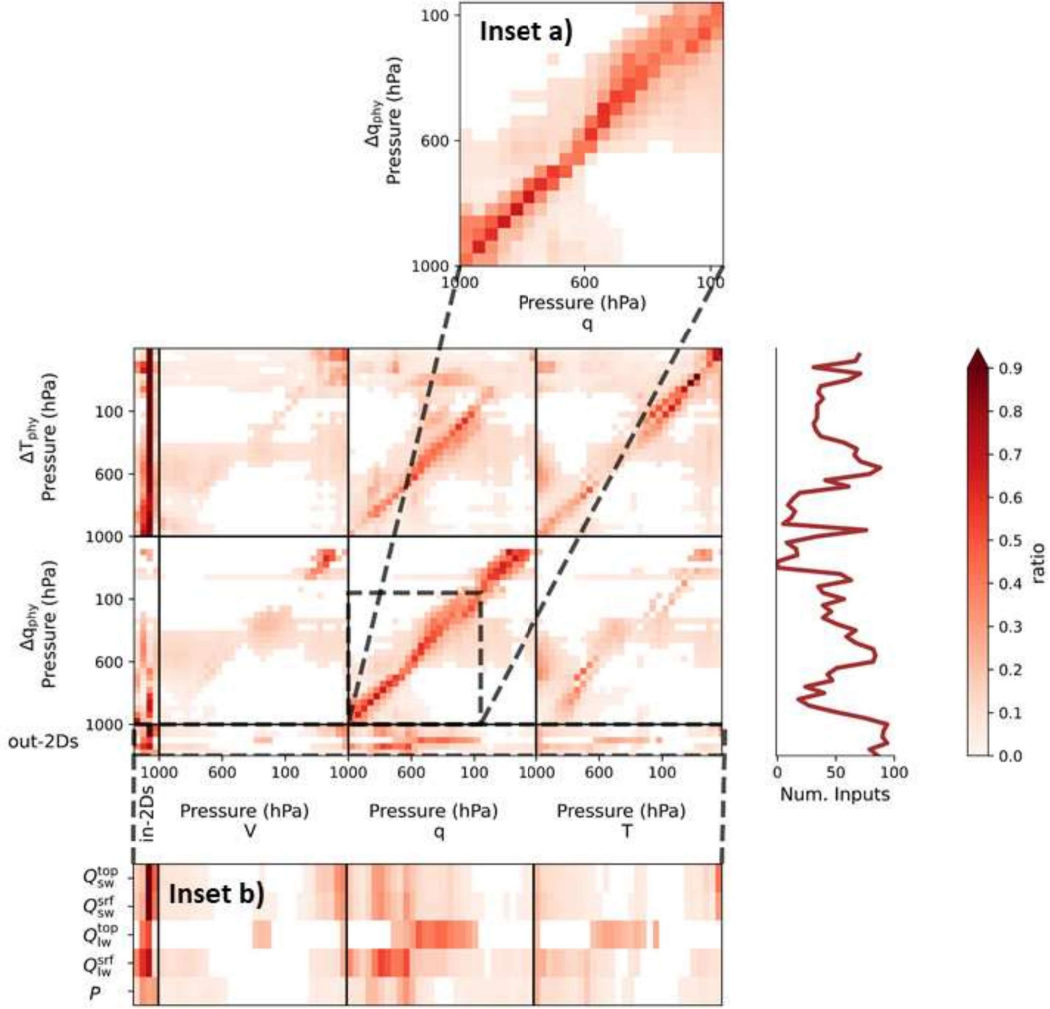

**Figure 2.** Causal discovery feature selection matrix of subgrid-scale processes in SPCAM for the varying optimized causal-threshold per output (see Supplementary Information). The inputs of the neural networks are given on the x-axes: 2–D variables ($Q_{lat}$, $Q_{sen}$, $Q_{sol}$, $P_{srf}$); and 3–D variables ($V$, $q$, $T$) from the surface ($10^3$ hPa) to the model's top (3 hPa), respectively. The outputs (subgrid-scale processes) are represented on the y-axes: 2–D variables ($P$, $Q_{lw}^{top}$, $Q_{sw}^{top}$, $Q_{lw}^{srf}$, and $Q_{sw}^{srf}$); and 3–D variables ($\Delta q_{phy}$ and $\Delta T_{phy}$) from the surface to the model's top, respectively. Insets (**a**) and (**b**) zoom-in into $\Delta q_{phy}$ and 2–D output variables, respectively. Contour colors represent the coefficient (absolute ratio) of the causal drivers appearance throughout the model's grid. Right panel shows the number of causal drivers for each output, with a mean number of inputs of 48 (51 % of the total).

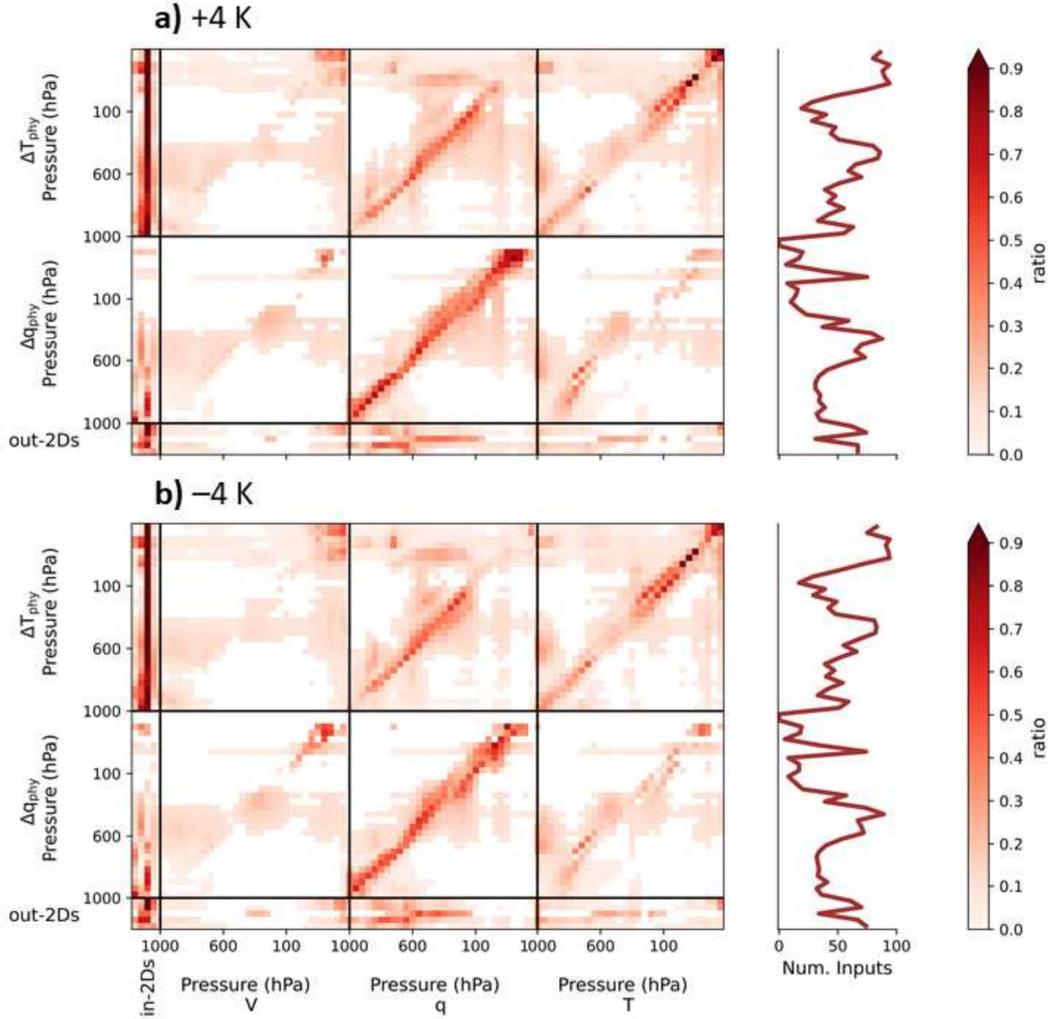

**Figure 3.** Same as Fig. 2, but for **a**) warming and **b**) cooling of global 4 K sea surface temperatures by adding a wavenumber one perturbation to the reference sea surface temperatures in increments of 1 K. Right panels show the number of causal drivers for each output. Both cases have an equivalent mean number of inputs of 48 (51 % of the total) compared to the reference climate (+0 K).

### 3.2 Mean Climate and Variability

We here couple both the standard NN emulation (Non-causalNN) and the new causally-informed NN emulation (Causally-informedNN) of SPCAM physics within CAM, result-ing in the Non-causalNNCAM and CausalNNCAM models respectively (Table 1). We evaluate their response when run online (i.e., coupled to the coarse-resolution model), using zonal-mean daily averaged output from 1 year prognostic runs with a 1 month spin-up (after reaching climate equilibrium; not shown). Note that instabilities generally would develop within the first 6 months (Ott et al., 2020).

The CausalNNCAM model accurately represents the mean tropospheric temperature and variability of the original model (Fig. 4) while the Non-causalNNCAM model shows significant cold biases in the tropics. Key features of the SPCAM simulations used here are the representation of a single ITCZ associated with a primary tropospheric tower of heating rates (warming; Fig. 4) and moistening rates (drying; Fig. S4) due to subgrid-scale processes, as well as secondary free-troposphere maxima at midlatitudes storm-tracks. While these features are well represented in CausalNNCAM, a spurious double-ITCZ is clearly represented in the Non-causalNNCAM. We note these biases are not present in the former non-causal hybrid model upon which this work builds, NNCAM (Rasp et al., 2018), which represents very well the mean climate of the original SPCAM simulation (see Section 4.2 and Fig. S11). It is also worth noting that the Causal$_{0.59}$NNCAM simulation shows similar deficiencies in the troposphere as in the non-causal realization. This suggests that a single optimized causal-threshold may well be too strict for a number of subgrid-scale processes (output predictands), which results in neglecting key physical drivers (input predictors) (see Fig. S5-S6). Stratospheric temperature biases are evident in all prognostic simulations with DL-based parameterization, and are very likely associated with the important role of ozone (missing variable as predictor in our NN setup) determining the climate in the stratosphere (WMO, 2018). These biases were also evident in NNCAM (Rasp et al., 2018).

The better ability of the causally-informed DL-based parameterization, CausalNNCAM, to realistically reproduce the mean ITCZ and characteristic midlatitude storm-track variability of the SPCAM reference simulation is also reflected in surface precipitation and net radiative fluxes (Fig. 5), as it accurately resolves the zonal patterns of precipitation and net radiation of SPCAM. In contrast, in Non-causalNNCAM precipitation is substantially underestimated, both mean and variability and a double-ITCZ pathology is evident. CausalNNCAM correctly captures precipitation peaks, though it is somewhat overestimated over the ITCZ and associated with stronger moistening rates (Fig. S4). Similarly, net radiative fluxes at the top of the atmosphere in Non-causalNNCAM are underestimated in the subtropics compared to SPCAM due to the double-ITCZ bias, deficiency largely overcome in CausalNNCAM. These results clearly show that CausalNNCAM not only reduces the dimensionality of the DL algorithm, which limits the impact of confounders such as the strong spatial (vertical and zonal) inter-correlations in the atmospheric profiles, but can also match the performance of NNCAM (Rasp et al., 2018). We reiterate that these biases in Non-causalNNCAM were not present in NNCAM (Fig. S11). Nevertheless, both non-causal and causal parameterizations accurately represent the physics of the test set (offline; Fig. S7-S8), and perform as well as the original NN (Rasp et al., 2018). Details about the disparities between offline and online performances are provided in Section 4.2. In principle, it would be possible to develop a good performing non-causal hybrid model by systematically training a very large number of NNs, as it has been already shown (Rasp et al., 2018).

Causal discovery helps improve DL-based learning of physical processes (i.e., parameterizations) by informing them with causal drivers. Two key open questions are whether such causally-informed neural networks: 1) lead to a reduced complexity of the system (lower dimensionality associated with greater input sparsity); and 2) can make more accurate predictions across climate regimes (improving generalization skills). To address the former question, we explore the performance of equivalent lower dimensional NNs compared to the causally-informed case (i.e., same number of inputs), but applying different methods to select the inputs. Both, randomly- and correlationally-informed NNs show worse performance compared to the causally-informed case for heating and moistening tendencies (Fig. S7). In addition, we use a linear version of the neural network parameterizations by replacing the activation functions with the identity function (i.e., removing non-linearity), to test whether inputs-to-outputs non-linearities are relaxed due to lower dimensionality in the causal case. The linear versions of the parameterizations show, as expected, a substantial drop in their performance compared to the non-linear versions (Fig. S7). Interestingly, however, the performance of the linear parameterizations for both, Non-causalNN and Causally-informedNN, are equivalent. Therefore, lower dimensionality alone explains little of the causally-informed NNs accurate performance, suggesting that it is largely related to the use of causal drivers (i.e., removing spurious links).

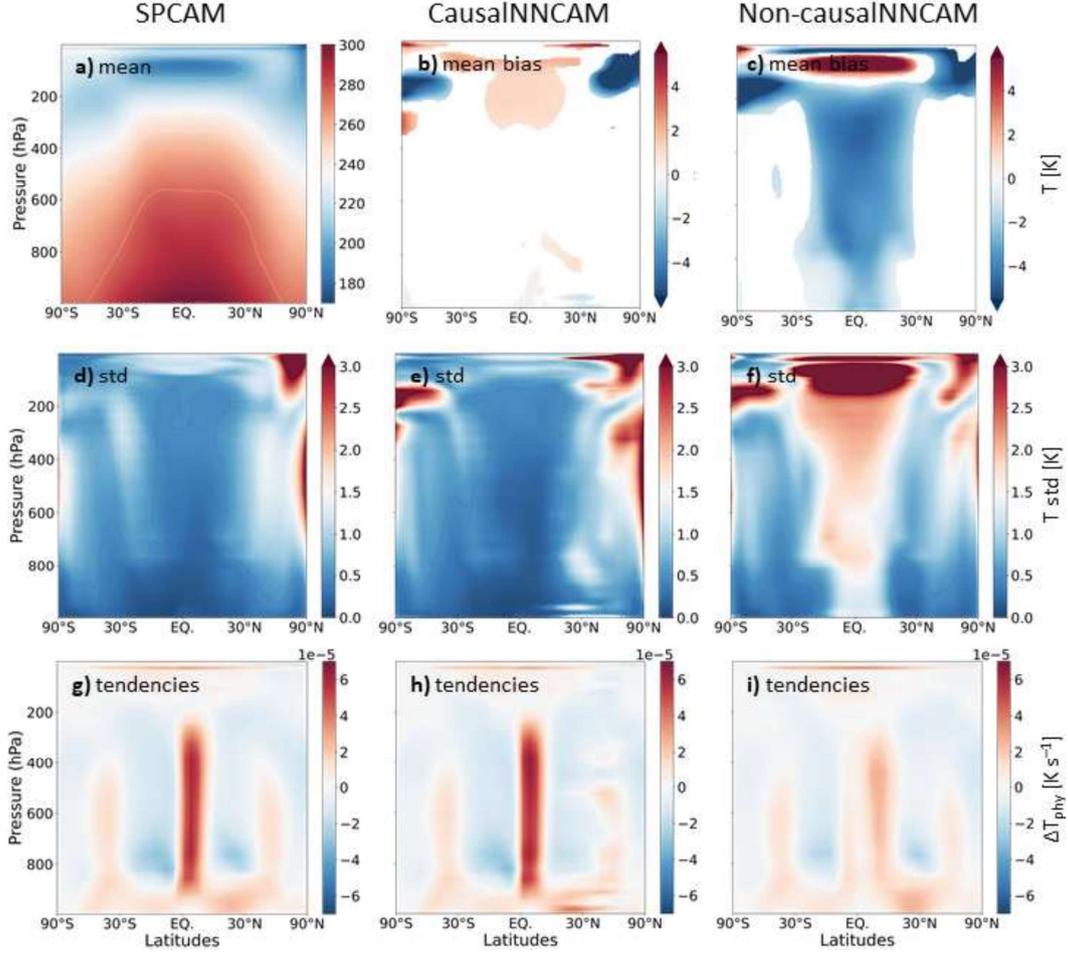

**Figure 4.** Zonal-mean climatologies of (**a-c**) temperature ($T$), (**d-f**) $T$ variability (standard deviation; std), and (**g-i**) heating tendencies ($\Delta T_{phy}$) for SPCAM, CausalNNCAM and Non-causalNNCAM. Contour colors for temperature biases (**b-c**) are for statistically significant differences at the 95 % confidence interval. Note Non-causalNNCAM biases are not present in NNCAM (Rasp et al., 2018), see Fig. S11.

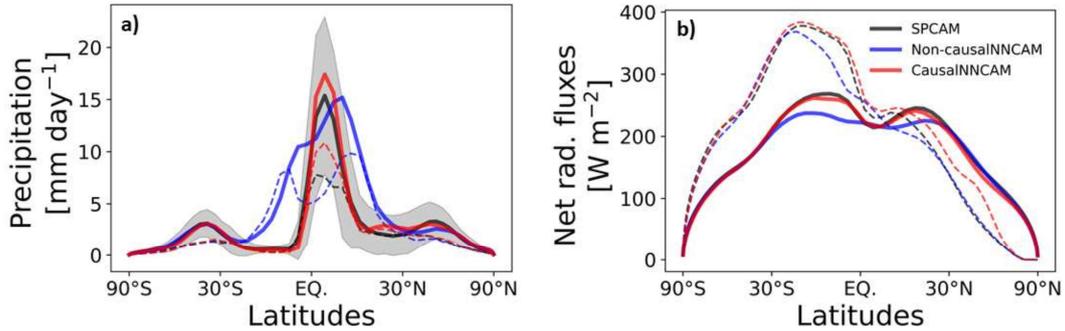

**Figure 5.** Zonal average climatologies of precipitation ($P$), and net radiative fluxes at the top of the atmosphere. **a**) Mean (thick solid lines) and standard deviation (thin dashed lines) are shown for $P$. Note the shaded grey area indicates the standard deviation around the mean of SP-CAM. **b**) Radiative longwave ($Q_{lw}^{top}$; solid lines) and shortwave ($Q_{sw}^{top}$; dashed lines) net fluxes are shown. Zonal mean values are area-weighted, i.e., cosine (latitudes). Note Non-causalNNCAM biases are not present in NNCAM (Rasp et al., 2018), see Fig. S11.

Then, we investigate the generalization capabilities of the DL-based parameterizations across ±4 K climates compared to the original climate model (see Supplementary Information). We find that causally-informed parameterizations retain similar generalization capabilities as the non-causal case, but without any substantial improvement with respect to the latter (Fig. S7). In particular, we find that our DL-based parameterizations, both Non-causalNN and Causally-informedNN, generalize poorly under the +4 K climate, which is in line with previous results (Rasp et al., 2018). DL algorithms usually optimize an objective using a training dataset. NNs make out-of-distribution predictions (extrapolation) such as across different climates, relying on implicit assumptions. There is no inherent guarantee that NNs will accurately generalize far beyond their training data (Beucler et al., 2021), even when using causal drivers. This extrapolation challenge leads to the failure of DL-based parameterizations when confronted with environmental conditions significantly different from their training data range (Rasp et al., 2018). Overall, these results suggest that while neural network parameterizations can be improved in combination with causality, the prediction skills across climates must be enhanced by other approaches (Beucler et al., 2021).

### 3.3 Neural Nets Explainability

Having demonstrated that causal discovery helps unveil direct physical drivers of subgrid-scale processes and that causally-informed prognostic simulation accurately represents the climate of the original SPCAM model, we turn here to explaining the pre-dictions of such DL-based parameterizations. Figure 6 shows the feature importance of unresolved processes predictions for both, Causally-informedNN and Non-CausalNN parameterizations, under the reference climate (+0 K), using a SHapley Additive exPlanations (SHAP) analysis (Lundberg & Lee, 2017; Shrikumar et al., 2017) (see Supplementary Information). In both cases, the predominant features are in line with physical understanding. Spurious features, however, are evident in the Non-causalNN case (Fig. 6a and Fig. S9). Heating and moistening rates in the lower troposphere are linked to temperature throughout the atmosphere. The absence of a clear pattern –but random links– is suggestive of non-physical spurious correlations. These spurious links are likely the result of the strong inputs-to-outputs inter-correlation vertically in the atmosphere due to convective processes. By construction, such spurious links are mainly missing in the causally-informed parameterization (Fig. 6b), which in turn shows stronger feature importance values for causal drivers compared to the latter (Fig. 6c).

## 4 Discussion

### 4.1 Causal-Threshold Optimization

Optimizing the causal-threshold to find robust and globally invariant causal drivers poses the unique challenge of lacking a ground truth causal graph, and therefore, relies on both expert knowledge and empirical performance. This work considers two threshold optimization cases: a single optimized quantile-threshold for all outputs (fixed value); and a varying quantile-threshold optimized for each output separately. While the single optimized quantile-threshold may appear more straightforward and "cleaner" (Fig. S1), it may fail to reflect the true complexity of the underlying causal relationships. In contrast, using a varying threshold optimized for each output introduces adaptability during the causal discovery phase and enables it to capture nuances in the data (Fig. 2). For example, mild causal relationships between heating tendencies in the upper stratosphere with environmental conditions throughout the atmosphere, present in the varying quantile-threshold, may be associated with structural artifacts of the original SPCAM model (e.g., models with top of the atmosphere below the stratopause present stratosphere-troposphere coupling issues) (Charlton-Perez et al., 2013). The superior empirical performance of the causal parameterization based on the varying threshold approach is associated with the set of causal drivers that better uncover hidden causal dependencies (Fig. 2), which may be overlooked by the more rigid single threshold strategy (Fig. S1).

### 4.2 Hybrid Model Stability and Performance

This work builds on a previous NN (architecture and hyperparameters) based on the same dataset, for which the resulting hybrid model (once the NN is coupled to the coarse climate model) ran stably and accurately represented the climate of the original SPCAM model (Rasp et al., 2018). We find a number of Causal$_q$NNCAM cases with suboptimal single optimized causal thresholds ($q \in [0.6, 0.8]$) that were unstable. Moreover, causally-informed parameterizations with the optimal causal threshold (quantile optimized; Table 1) but with simpler architectures (shallower and less complex) were also unstable once coupled to the host climate model. This result is in agreement with previous work (Rasp et al., 2018). Nevertheless, we find that Non-causalNNCAM, Causal$_{0.59}$NNCAM, and CausalNNCAM run stably without climate drifts (spurious and increasing long-term errors compared to SPCAM; not shown).

The DL-based parameterizations presented here (Non-causalNN, Causally-informed$_{0.59}$NN, and Causally-informedNN) perform as well as the original NN (Rasp et al., 2018) in the test set (offline; Fig. S7). Particularly, we note that Non-causalNN (and Causally-informed$_{0.59}$NN; not shown) accurately captures both the ITCZ and midlatitudes storm-tracks as represented by SPCAM (Fig. S8 top row). We find that DL-based parameterizations for surface precipitation and net radiative fluxes following a better architecture found by a sys-tematic hyperparameter tuning using an analogous SPCAM dataset (Hertel et al., 2020), lead to a marginal or negligible performance improvement (Fig. S8 bottom row). Notably, there are disparities between offline and online performances (Brenowitz, Henn, et al., 2020). In theory, it would be possible to develop equally good, or even better, performing hybrid models compared to the original NNCAM model (Rasp et al., 2018) if one trains a large number of NNs (Ott et al., 2020; Lin et al., 2023). In spite of that, this work advances DL-based parameterizations for climate models by implementing causal discovery, and demonstrates these newly developed causally-informed NNs better respect the underlying physical processes, with improved interpretability and without compromising performance skills (offline and online).

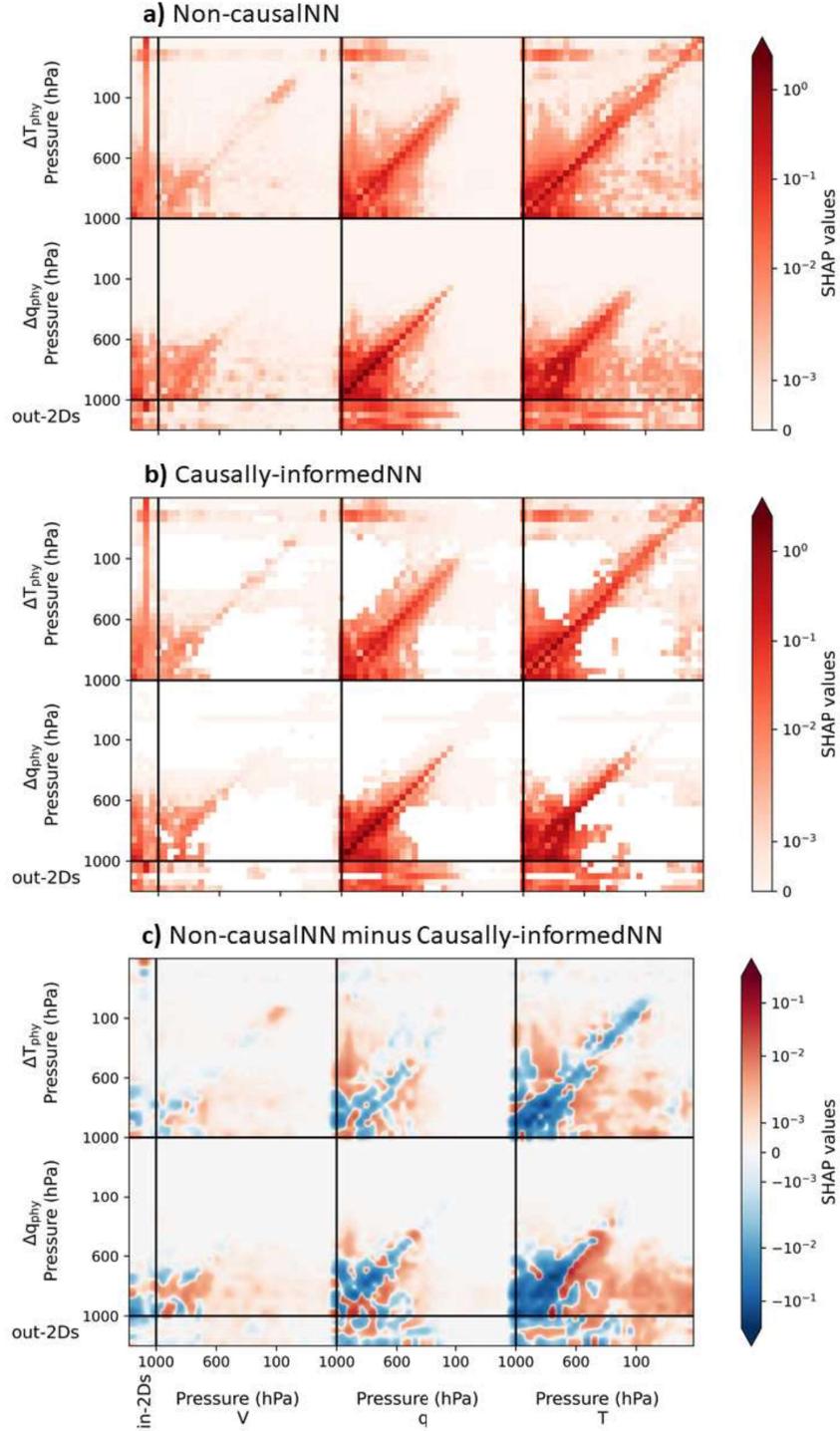

**Figure 6.** Feature importance mapping of subgrid-scale processes predictions. Explanations are based on absolute averaged values of the SHapley Additive exPlanations (SHAP; see Supplementary Information) (Lundberg & Lee, 2017; Shrikumar et al., 2017) for: (**a**) Non-causalNN; (**b**) Causally-informedNN and (**c**) the difference between both parameterizations (smoothed via gaussian interpolation). Same as in Fig 2, the inputs of the neural networks are given on the x-axes, and the outputs (subgrid-scale processes) are represented on the y-axes. 3–D variables are shown from the surface ($10^3$ hPa) to the model's top (3 hPa). Absolute averaged SHAP values are shown in symlog scale, and are calculated for over 4000 random samples of the test set compared to the train set (as background).

## 5 Conclusions

Data-driven parameterizations of subgrid-scale processes based on SRMs are able to represent to a good extent the climate of the original simulation once coupled to the coarser climate model (i.e., hybrid model) (Rasp et al., 2018; Yuval & O'Gorman, 2020; Watt-Meyer et al., 2021; Bretherton et al., 2022). Hybrid models can potentially alleviate persistent biases in coarse climate model simulations, and improve future climate projections. However, instabilities in hybrid models have been difficult to overcome and prognostic skills are challenging even in idealized simulations (e.g., aquaplanet settings without topography). It may well be that the sources of such instabilities and prognostic skills are associated with spurious non-physical relationships learned by the ML algorithm due to strong vertical inter-correlations, as well as fitting to noise. Current approaches to achieve stable hybrid models fail to care for the causes (e.g., deepening the deep learning algorithm or ablating the stratosphere) (Rasp et al., 2018; Brenowitz & Bretherton, 2019). An approach that is scalable and can reliably target the causes to overcome these issues would be a key breakthrough for ML-based parameterizations.

Here, we present a novel approach that combines causal discovery and DL to improve climate models and projections. We demonstrate that causal discovery robustly unveils causal (physical) drivers of subgrid-scale processes across different climate regimes, while improving interpretability and trust in the DL algorithm. Our causally-informed data-driven model also runs stably when coupled to the host coarse resolution model and generates a climate (mean and variability) close to the original simulation under the reference climate (within the distribution of the training dataset). We showed that causally-informed NNs prevent obvious spurious links in conventional DL-based parameterizations, leading to greater attention of the algorithm to the physical drivers compared to the latter.

Causal discovery, however, requires expert knowledge. In particular, we provided a solution to optimizing the causal-threshold (i.e., significance of the causal drivers), by running statistics over a number of causal graphs and testing the performance of the related causally-informed NNs. Yet, we avoided a systematic hyperparameter tuning of the original DL algorithm (Rasp et al., 2018) to find a stable and skillful performing hybrid model (Hertel et al., 2020). Moreover, we demonstrate that causal discovery, both $PC_1$ and linear Lasso regression, can identify key causal drivers of subgrid-scale processes that respect the underlying physical mechanisms (i.e., removing redundant information and non-physical links), for which standard feature selection methods, such as linear correlation, clearly fail. Future work will test this approach in more challenging and realistic setups (e.g., historical simulations with varying forcings and a real topography), as well as extend this method to integrate causality with state-of-the-art advances in deep learning approaches (Camps-Valls et al., 2021).

This work presents a fundamental and novel step in overcoming major challenges of data-driven models of physical processes (e.g., in parameterizations for climate models), paving the way towards improving climate models and projections via causally-based machine learning techniques. Explicitly using direct drivers in deep learning methods to represent physical processes is a key challenge that our methodology addresses, which in turn helps solve the problem of finding more reliable and reproducible data-driven parameterizations. Furthermore, advances in machine learning techniques are rapidly offering potential solutions to other limitations, such as generalization capabilities. The combination of causal discovery and deep learning presented here introduces a powerful new approach that opens a new window into process-based representation of complex processes not only for Earth system science but also in other scientific disciplines.

## Data Availability Statement

The code used to train the neural networks and to produce all figures of this manuscript is archived on Zenodo: Software - (Solino & Iglesias-Suarez, 2023). An example of SPCAM data is also archived on Zenodo: Data - (Rasp, 2019).

## Author Contributions

F.I.-S. developed the causally-informed neural network parameterizations and performed the analysis with support of B.S.-F., and with contributions from all co-authors he conceptualized the research. V.E. and P.G. formulated the research question and concept, J.R. framed the causal approach, T.B. pre-processed the original SPCAM simulations, and F.I.-S. performed the causally-informed neural network climate simulations with support of M.P. All authors discussed the methodology and findings. F.I.-S. wrote the manuscript with contributions from all authors.

## Additional Information

The authors declare no competing interests. The online version contains Supplementary Information, including detailed description of the SPCAM model setup, neural networks specifications, and causal discovery optimization.

## Acknowledgments
Funding for this study was provided by the European Research Council (ERC) Synergy Grant "Understanding and modeling the Earth System with Machine Learning (USMILE)" under the Horizon 2020 research and innovation programme (Grant agreement No. 855187). F.I.-S. is a postdoc of the ELLIS Postdoc Program and acknowledges travel support from the European Union's Horizon 2020 research and innovation programme under ELISE (Grant Agreement No. 951847). Additionally, T.B. acknowledges funding from the Columbia University sub-award 1 (PG010560-01). P.G. and M.P. acknowledge funding from the National Science Foundation Science and Technology Center, Learning the Earth with Artificial intelligence and Physics, LEAP (Grant number 2019625). M.P. acknowledges funding from the US Department of Energy Advanced Scientific Computing Research program (DE-SC0022331). J.R. has received funding from the European Research Council (ERC) Starting Grant CausalEarth under the European Union's Horizon 2020 research and innovation program (Grant Agreement No. 948112). This work used resources of both, the Deutsches Klimarechenzentrum (DKRZ) granted by its Scientific Steering Committee (WLA) under project ID 1179 (USMILE), and the supercomputer JUWELS at the Jülich Supercomputing Centre (JSC) under the Earth System Modelling Project (ESM).

# Supporting Information for "Causally-informed deep learning to improve climate models and projections"

Fernando Iglesias-Suarez[1], Pierre Gentine[2,3], Breixo Solino-Fernandez[1], Tom Beucler[4], Michael Pritchard[5,6], Jakob Runge[7,8], and Veronika Eyring[1,9]

[1]Deutsches Zentrum für Luft- und Raumfahrt e.V. (DLR), Institute of Atmospheric Physics, Oberpfaffenhofen, Germany

[2]Department of Earth and Environmental Engineering, Center for Learning the Earth with Artificial intelligence and Physics (LEAP), Columbia University, New York, USA

[3]Earth and Environmental Engineering, Earth and Environmental Sciences, Learning the Earth with Artificial intelligence and Physics (LEAP) Science and Technology Center, Columbia University, New York, USA

[4]University of Lausanne, Institute of Earth Surface Dynamics, Lausanne, Switzerland

[5]University of California, Department of Earth System Science, Irvine, USA

[6]NVIDIA Corporation, Santa Clara, USA

[7]Deutsches Zentrum für Luft- und Raumfahrt e.V. (DLR), Institute of Data Science, Jena, Germany

[8]Technische Universität Berlin, Institute of Computer Engineering and Microelectronics, Berlin, Germany

[9]University of Bremen, Institute of Environmental Physics (IUP), Bremen, Germany

**Contents of this file**
1. Text S1 to S4
2. Table S1
3. Figures S1 to S11

**Introduction** This text complements the description of the concept of this study, the SPCAM model setup, neural networks specifications, causal discovery optimization, and neural nets explainability provided on the main manuscript.

**Text S1. Superparameterized Community Atmosphere Model v3.0 (SPCAM)** The superparameterization component (SP) is spun-up at the beginning of the simulation and subcycles every 20 seconds given the large-scale tendencies (Collins et al., 2006). At the end of every time step (30 minutes), the horizontal mean of state variables for temperature, moisture and condensate from the SP component update the resolved fields in its host (Benedict & Randall, 2009). Note that unlike traditional parameterizations, the SP component runs continuously throughout the simulation after the initial spin-up, which adds a "memory" effect to the subgrid-scale processes affecting the large-scale tendencies. The memory of subgrid-scale processes is not explicitly treated during the training of the neural networks (NNs); rather we assume its importance is secondary (Jones et al., 2019). After the SP component, the radiation scheme is called using the explicitly resolved vertical distribution of clouds among the large-scale resolved fields. Finally, surface fluxes are computed following a simple bulk scheme using the host model's coarse state fields, and then the dynamical core.

Solar insolation follows a diurnal cycle in perpetual Southern Hemisphere solstice. Sea surface temperatures (SSTs) are imposed following a zonally symmetric setup but with a shift in maximum temperatures five degrees North of the equator (Andersen & Kuang, 2012):

$$SST(\phi) = 2 + \frac{27}{2}(2 - \zeta - \zeta^2), \quad (1)$$

with $SSTs$ in Celcius, latitudes ($\phi$) in degrees, and

$$\zeta = \begin{cases} sin^2(\pi\frac{\phi-5}{110}) & 5 < \phi \leq 60 \\ sin^2(\pi\frac{\phi-5}{130}) & -60 \leq \phi < 5 \\ 1 & \text{if } |\phi| < 60 \end{cases}. \quad (2)$$

Sensitivity experiments comprise global changes in SSTs ($\pm 4$ K) by adding a wavenumber one perturbation to the reference ($+0$ K) SSTs in increments of 1 K:

$$SST'(\lambda, \phi) = 3\,cos\left(\frac{\lambda\pi}{180}\right) cos\left(0.5\pi \frac{\frac{\phi\pi}{180} - 5}{30}\right)^2$$
$$\text{if } -25 \leq \phi \leq 35, \quad (3)$$

with longitudes ($\lambda$) in degrees. The SPCAM model source code used here, including the neural network implementation via the Fortran-Keras Bridge (Ott et al., 2020), is available at https://gitlab.com/mspritch/spcam3.0-neural-net (causalcoupler branch; commit hash: 5ebff0a6). Further details of the SPCAM model are provided elsewhere (Khairoutdinov & Randall, 2001; Collins et al., 2006; Pritchard et al., 2014; Pritchard & Bretherton, 2014).

**Text S2. Neural network setup** Building on previous hybrid modeling work (Rasp et al., 2018), we develop and train all NNs used here with Keras (https://keras.io/) built on top of Tensorflow 2 (https://www.tensorflow.org/). All NNs were trained for 18 epochs using: a batch size of 1,024; the LeakyReLU activation function set to $max(0.3x, x)$; the Adam optimizer (Kingma & Ba, 2014); a starting learning rate of $1 \times 10^3$ subsequently divided by 5 every 3 epochs; and a mean squared error loss function.

Each input field was normalized by subtracting its mean across samples and then dividing it by its maximum range, whereas the outputs were normalized to bring them to the same order of magnitude (Table S1).

The computational costs of CausalNNCAM and Non-causalNNCAM are virtually the same. However, these hybrid models, consisting of 65 single-output NNs coupled to the host model via the Fortran-Keras Bridge, result in similar computational costs compared to SPCAM. Therefore, CausalNNCAM and Non-causalNNCAM are approximately 20 times slower than a former hybrid model consisting in one multi-output NN (Rasp et al., 2018). Nevertheless, we emphasize that our study is a proof of concept focused on the added value of combining causality and ML, and envisage an overall speed up of the hybrid models could be attained by developing a method that enables multi-output causally-informed NNs, as well as advanced software engineering solutions and coding (e.g., efficiently coupling machine learning algorithms with Fortran code).

**Text S3. Causal drivers optimization** We optimize the causal threshold during the causal discovery phase (i.e., finding the most important causal drivers of the subgrid-scale processes in SPCAM). The causal drivers determine the input layer of the causally-informed NNs, while the rest of the NNs setup remains unchanged (see above). Two optimization approaches are considered: 1) a single optimized causal threshold for all outputs; and 2) a varying optimized causal threshold for each output.

The single optimized causal threshold approach focuses on $\Delta T_{phy}$ and $\Delta q_{phy}$ at the level closest to the surface (992 hPa), where the causally-informed NNs show worst offline performance compared to the non-causal NNs (reference case). The single optimized threshold is chosen following two conditions:

1. $R^2_{causally-informedNN(thr)} \geq R^2_{non-causalNN} \leftrightarrow R^2_{non-causalNN} > 0$
2. $\max(thr)$.

The NNs were trained using the reference simulation (+0 K), and their performance was computed applying the coefficient of determination ($R^2$) to the test sets of all simulations considered ($-4$ K, $+0$ K and $+4$ K). While the first condition selects those causally-informed NNs that perform at least as good as its non-causal NN counterpart (as long as it shows some skill), the second condition selects the causally-informed NN with the most stringent threshold (minimum number of causal drivers).

There are two causal threshold definitions considered (Fig. S1):

$$causal'(Y_t^j) = X_{t-\tau}^i : P(X_{t-\tau}^i \to Y_t^j \in causal_g(Y_t^j)) > quantile \quad (4)$$

and,

$$causal'(Y_t^j) = X_{t-\tau}^i : \left(\frac{\#(X_{t-\tau}^i \in causal_g(Y_t^j))}{N_g}\right) > ratio. \quad (5)$$

The quantile-based definition considers causal drivers of each output ($causal'(Y_t^j)$), the inputs ($X_{t-\tau}^i$) for which their probability of being causally-linked to the given output is greater than a certain quantile (threshold). The ratio-based definition considers causal drivers of each output, the inputs for which their ratio of being causally-linked to the given output across the model's grid ($N_g = 8,192$) is greater than a certain value (threshold). Figure S10 shows the $R^2$ value for all causally-informed NNs explored compared to the non-causal NNs for both, $\Delta T_{phy}$ and $\Delta q_{phy}$. While using the definition of the ratio-based threshold no causally-informed NN explored met the above conditions, we find the optimal 0.59 value for the quantile-based approach (See Table 1; Causal-threshold: quantile 0.59).

Using the quantile-based threshold definition, a varying optimized causal threshold for each output is achieved using the Grid Search algorithm of the SHERPA package (Hertel et al., 2020). We explore a threshold range of 0.-0.95 at intervals of 0.05. For each output, 20 NNs were trained –one per threshold step– using the reference simulation (+0 K), and the optimal threshold is based on the minimum mean squared error of the validation set (6 floating points).

Figure S7 shows the offline performance ($R^2$) of the Causally-informed$_{0.59}$NN (single optimized causal-threshold) and Causally-informedNN (varying optimized causal-threshold) cases for $\Delta T_{phy}$ and $\Delta q_{phy}$, using the test set of the reference simulation (+0 K). Although Causally-informed$_{0.59}$NN shows similar offline performance compared to Causally-informedNN, its hybrid model (Causal$_{0.59}$NNCAM) shows a double Intertropical Convergence Zone bias (see discussion in main text and Fig. S5 and S6).

**Text S4. Neural nets explainability** We use the SHapley Additive exPlanations (SHAP) (Lundberg & Lee, 2017) game theoretic approach to explain the predictions of subgrid-scale processes by the NNs. Although the computation of the exact Shapley value is challenging, there are different SHAP built-on methods to approximate it. The algorithm used here is DeepExplainer, specifically tailored for deep learning models, which is based on the Deep Learning Important FeaTures (DeepLIFT) (Shrikumar et al., 2017). DeepExplainer decomposes the prediction (output) of a neural network on the different inputs. This is achieved by approximating the difference of the output from a distribution of background outputs with regard to the difference of the input from a distribution of background inputs. The complexity of the method scales linearly with the number of background samples ($n$), and the variance of the expectation estimates scales by approximately $1/\sqrt{n}$. Although 1000 samples would give already a very good approximation to the exact Shapley value, we use 4096 random samples across 1440 time-steps (∼1 month) and the full horizontal model grid (8192 points; latitude by longitude).

**Table S1.** Summary of neural networks inputs and output fields.

| Inputs | Units | Outputs | Units | Normalization |
|---|---|---|---|---|
| Temperature, $T(p)$ | K | Temperature tendencies, $\Delta T_{phy}(p)$ | $Ks^{-1}$ | $C_p$ |
| Specific humidity, $q(p)$ | $kgkg^{-1}$ | Moistening tendencies $\Delta q_{phy}(p)$ | $kgkg^{-1}s^{-1}$ | $L_v$ |
| Meridional wind, $V(p)$ | $ms^{-1}$ | Net shortwave radiative heat flux at TOA, $Q_{sw}^{top}$ | $Wm^{-2}$ | $10^{-3}$ |
| Surface pressure, $P_{srf}$ | Pa | Net longwave radiative heat flux at TOA, $Q_{lw}^{top}$ | $Wm^{-2}$ | $10^{-3}$ |
| Incoming solar radiation, $Q_{sol}$ | $Wm^{-2}$ | Net shortwave radiative heat flux at the surface, $Q_{sw}^{srf}$ | $Wm^{-2}$ | $10^{-3}$ |
| Sensible heat flux, $Q_{sen}$ | $Wm^{-2}$ | Net longwave radiative heat flux at the surface, $Q_{lw}^{srf}$ | $Wm^{-2}$ | $10^{-3}$ |
| Latent heat flux, $Q_{lat}$ | $Wm^{-2}$ | Precipitation, P | $kgm^{-2}d^{-1}$ | $1.728\times10^6$ |

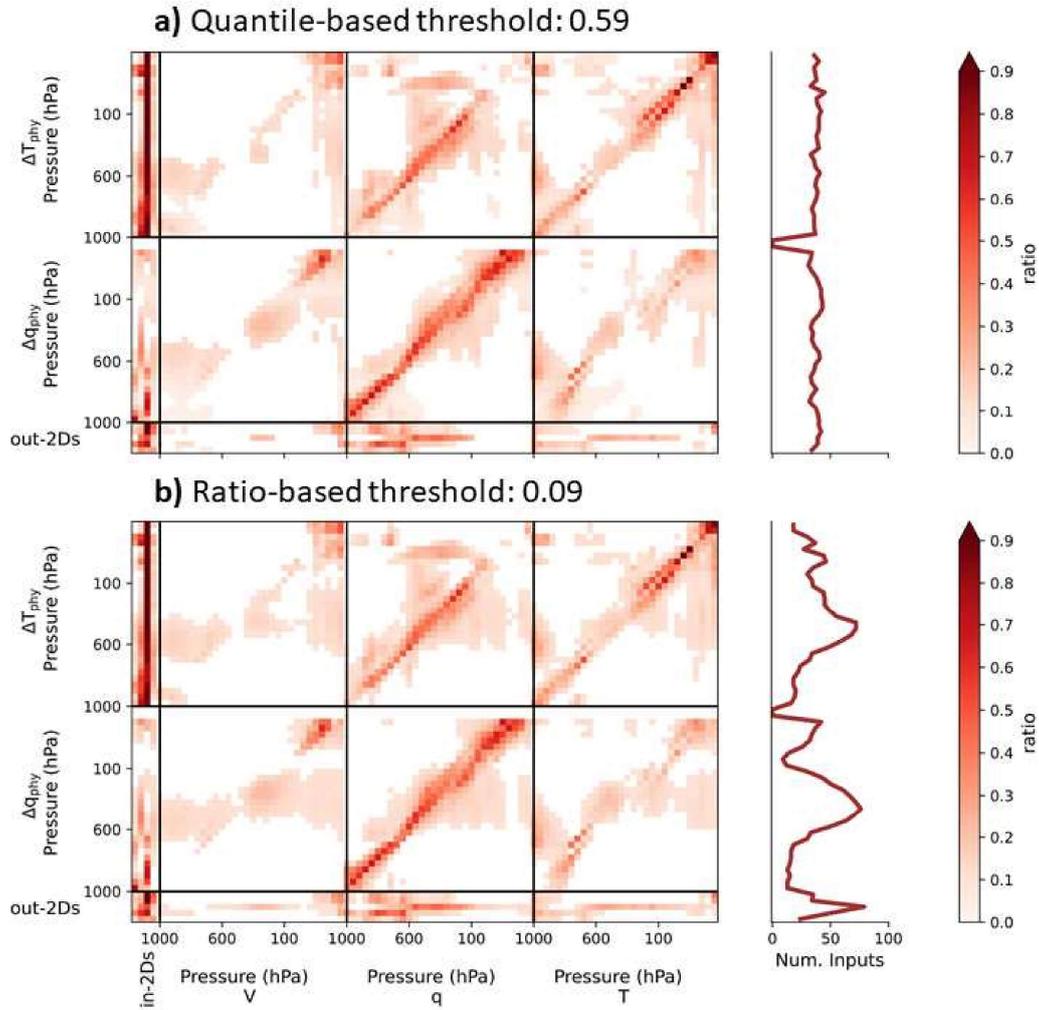

**Figure S1.** Same as Fig. 2, but for single optimized **a**) quantile-based threshold (0.59), and **b**) ratio-based threshold (0.09). Right panels show the number of causal drivers for each output, with a mean number of inputs of 36 (39 % of the total) and 35 (38 % of the total) for the quantile-based-threshold of 0.59 and the ratio-based threshold of 0.09, respectively.

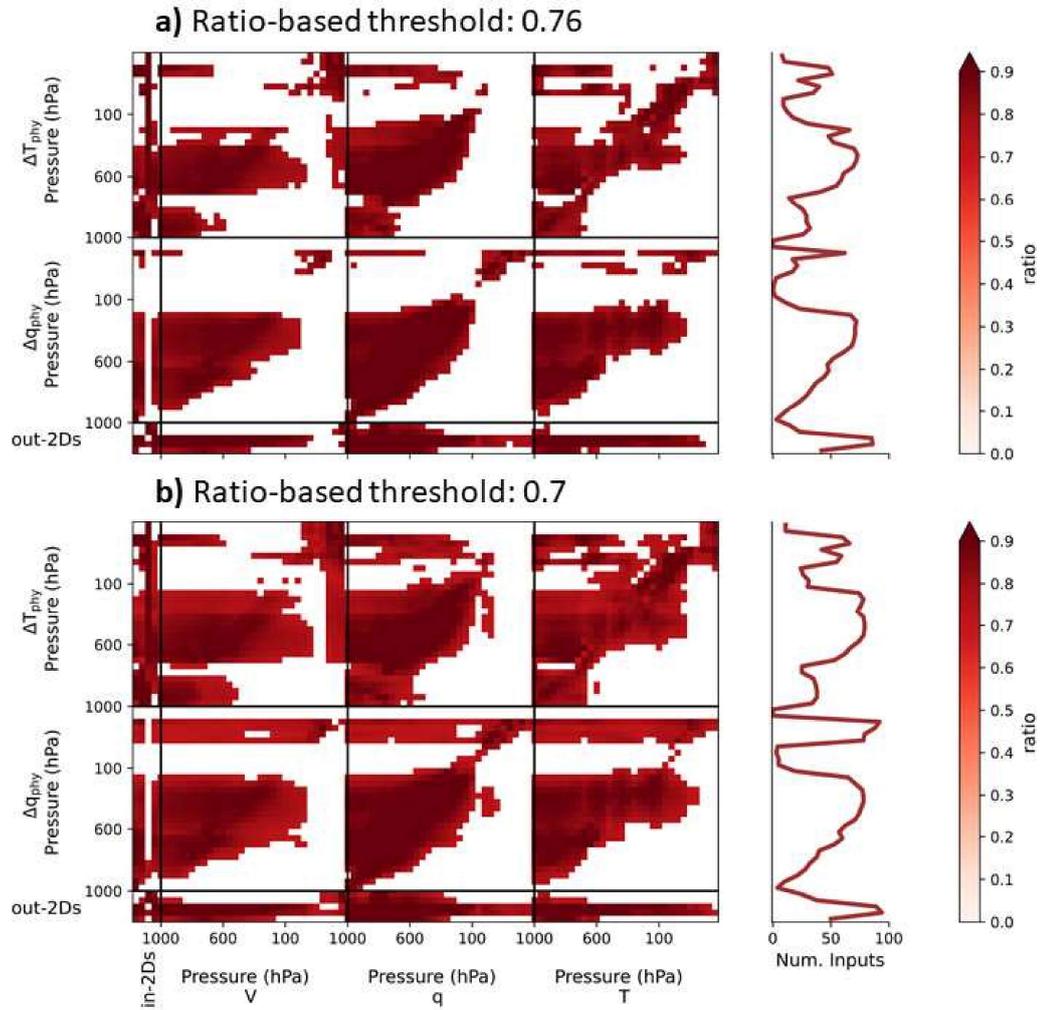

**Figure S2.** Same as Fig. S1, but for Pearson correlation ratio-based thresholds of **a**) 0.76, and **b**) 0.7. Right panels show the number of causal drivers for each output, with a mean number of inputs of 36 (39 % of the total) and 48 (51 % of the total) for the ratio-based thresholds of 0.76 and 0.7, respectively.

X - 6 :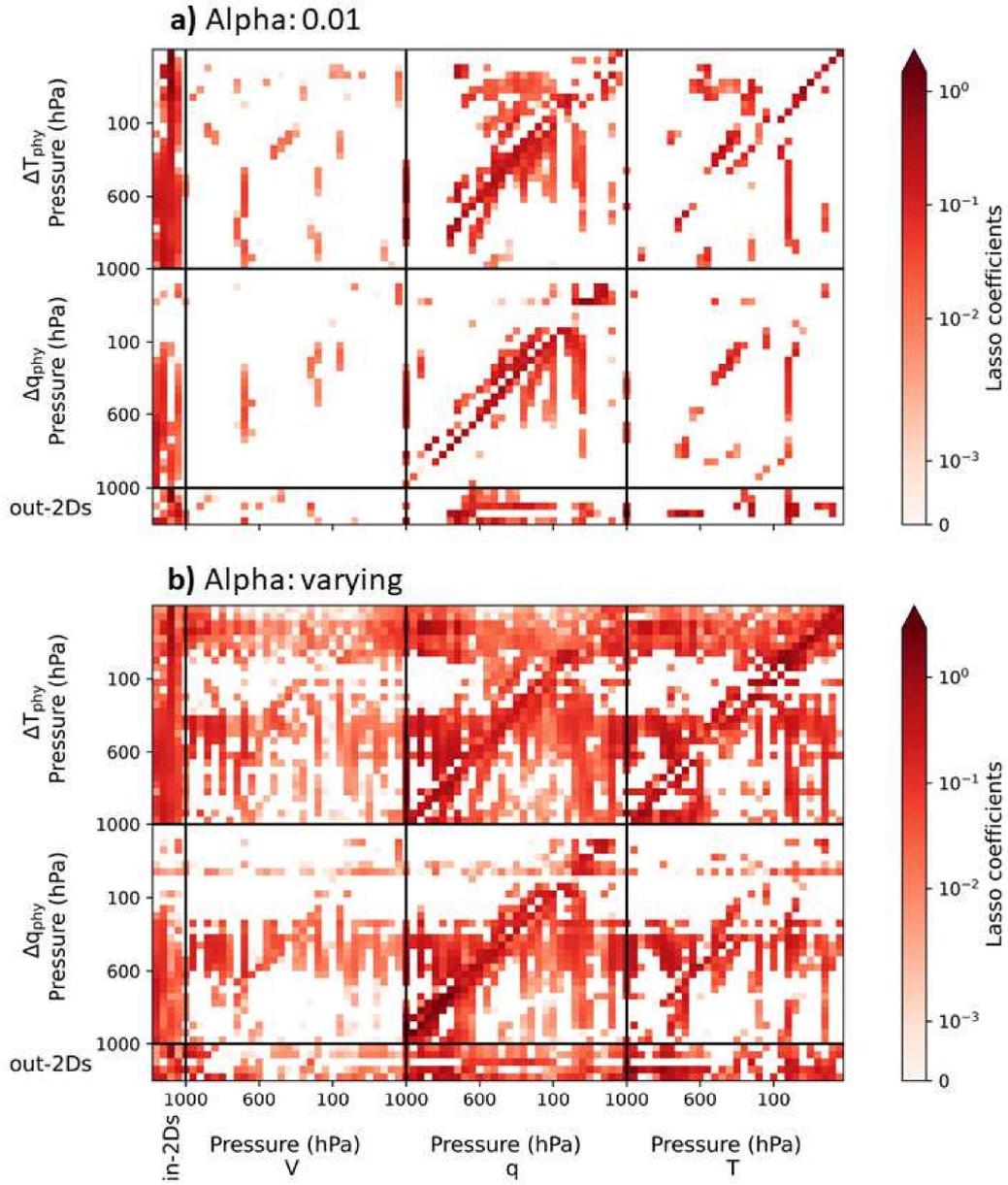

**Figure S3.** Same as Fig. S1, but for Lasso regression with **a**) 0.01 alpha and **b**) varying alpha. Note varying alpha in **b**) is chosen to obtain a similar number of inputs as in the quantile-optimized causal-threshold case (Fig. 2).

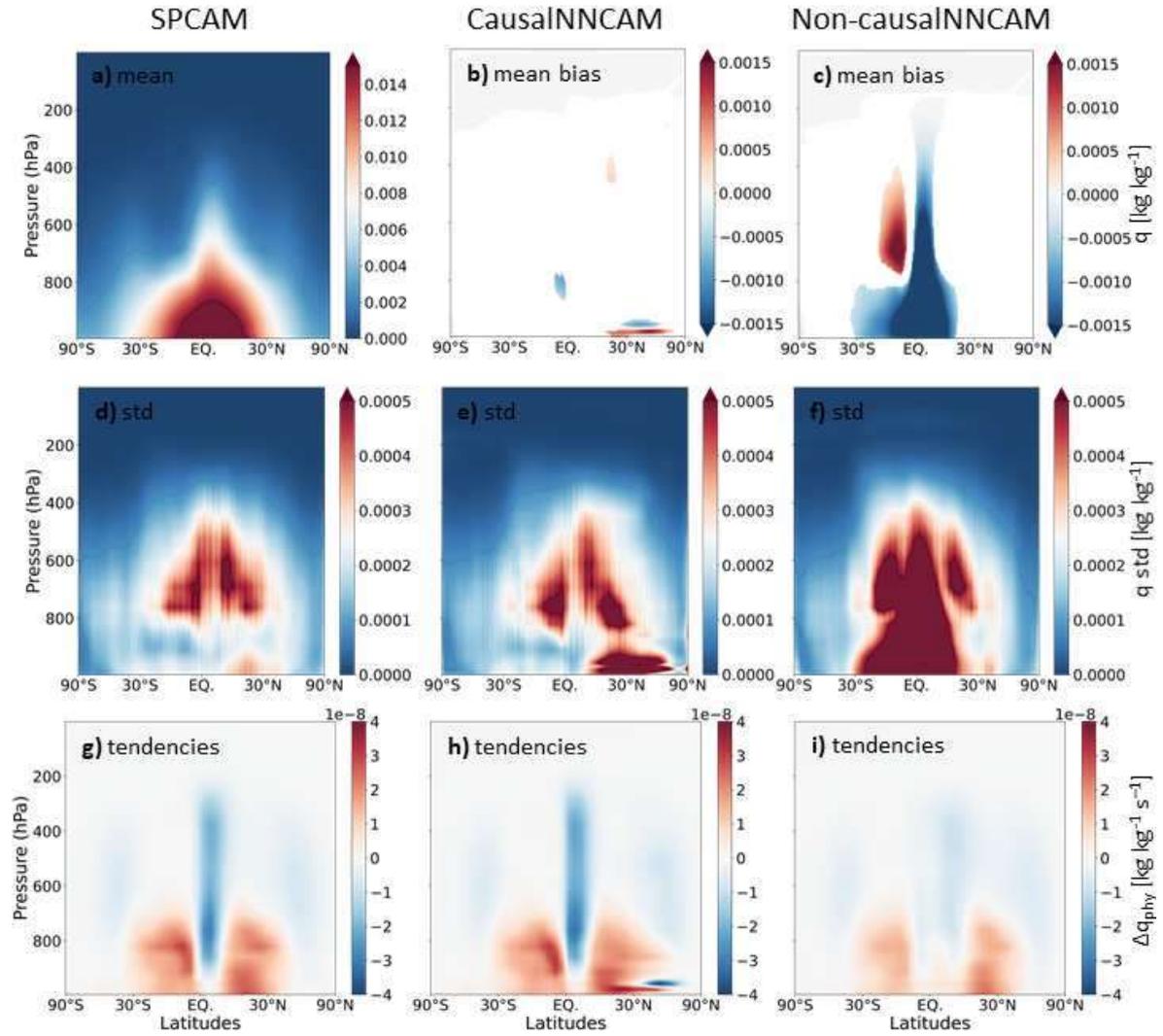

**Figure S4.** Same as Fig. 4, but for specific humidity ($q$) and moistening rates ($\Delta q_{phy}$).

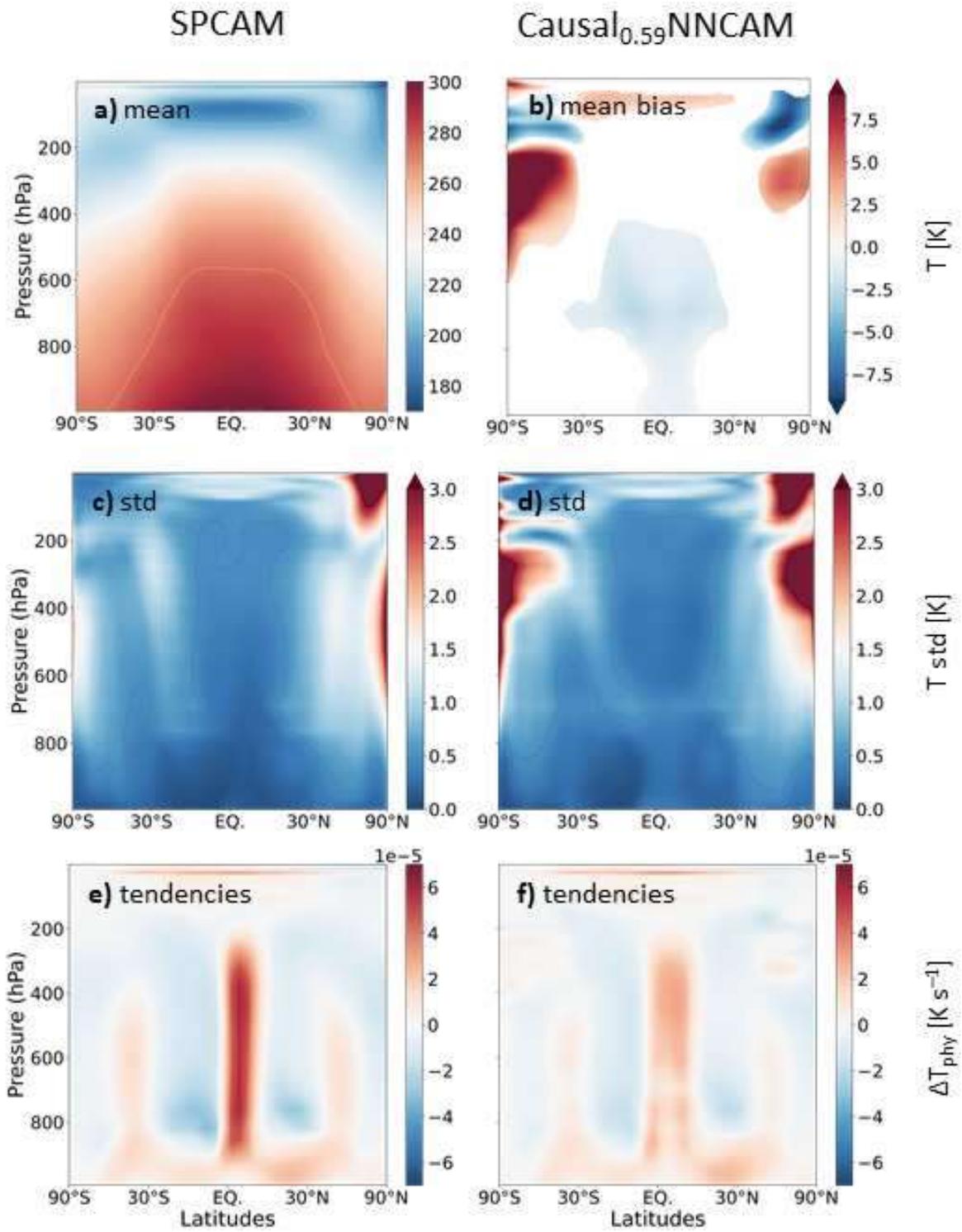

**Figure S5.** Same as Fig. 4, but for Causal$_{0.59}$NNCAM.

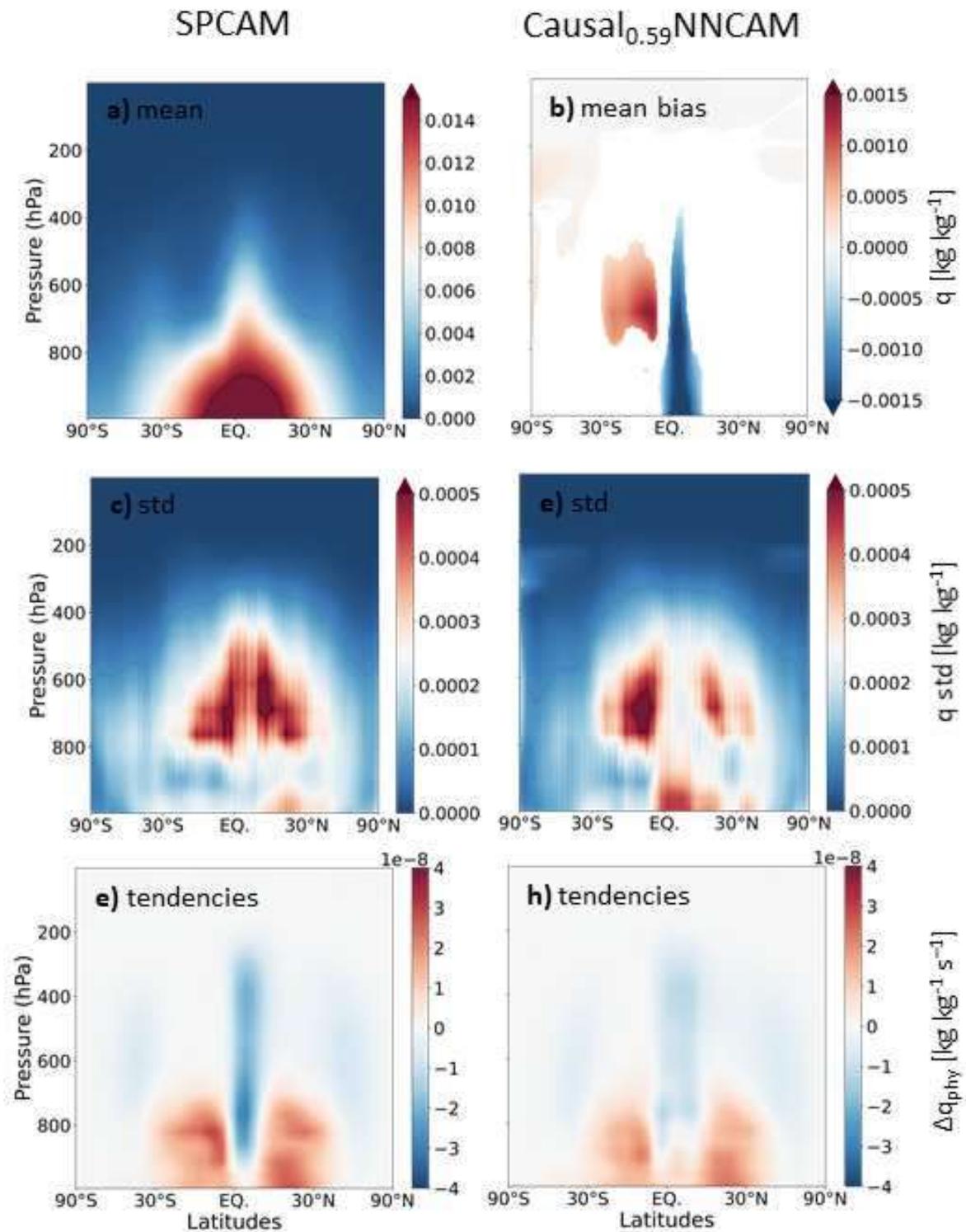

**Figure S6.** Same as Fig. S4, but for Causal$_{0.59}$NNCAM.

xy

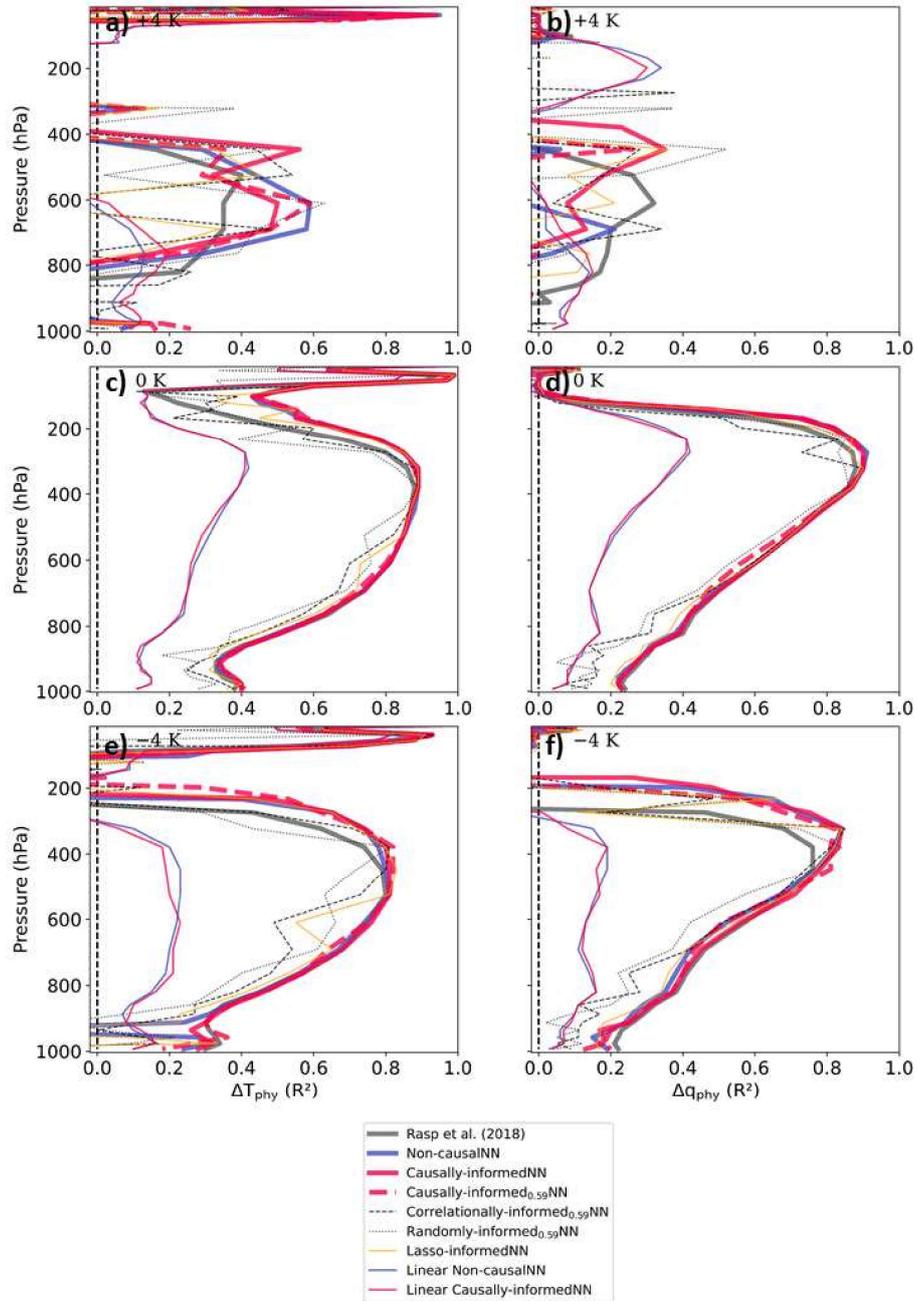

**Figure S7.** Vertically resolved coefficient of determination ($R^2$), averaged horizontally and in time, for heating rates ($\Delta T_{phy}$) and moistening rates ($\Delta q_{phy}$) of neural network (NN) parameterizations trained on the reference climate (+0 K). $R^2$ is calculated using the test sets of each SPCAM simulation case ($-4$ K, $+0$ K and $+4$ K). Linear version of the neural network parameterizations (thin solid lines) are for the activation identity functions. Correlationally-informed (dashed black lines) and randomly-informed (dotted black lines) NNs use the same number of inputs as in the single optimized causal-threshold ($q=0.59$) case. Lasso-informed (solid yellow line) NN use similar number of inputs as in the quantile-optimized causal-threshold case (Fig. S3b).

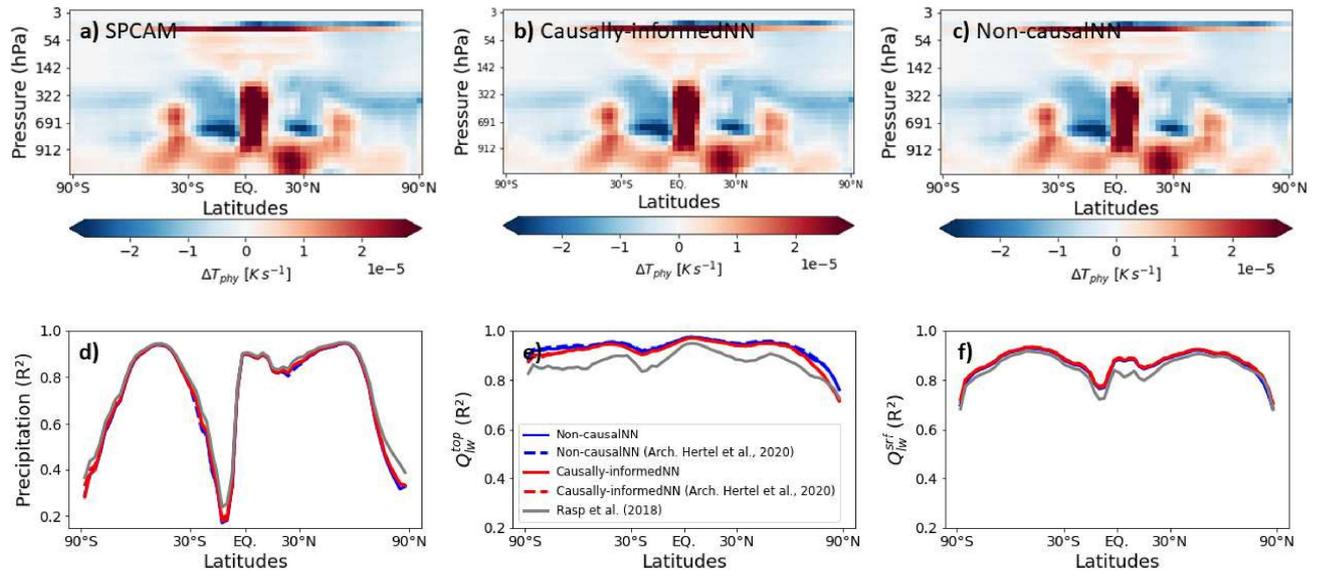

**Figure S8.** (**top row**) Zonal-mean climatologies of heating tendencies ($\Delta T_{phy}$), and (**bottom row**) latitudinally resolved coefficient of determination ($R^2$) of surface precipitation ($P$) and net longwave radiative heat fluxes ($Q_{lw}^{top}$, $Q_{lw}^{srf}$).

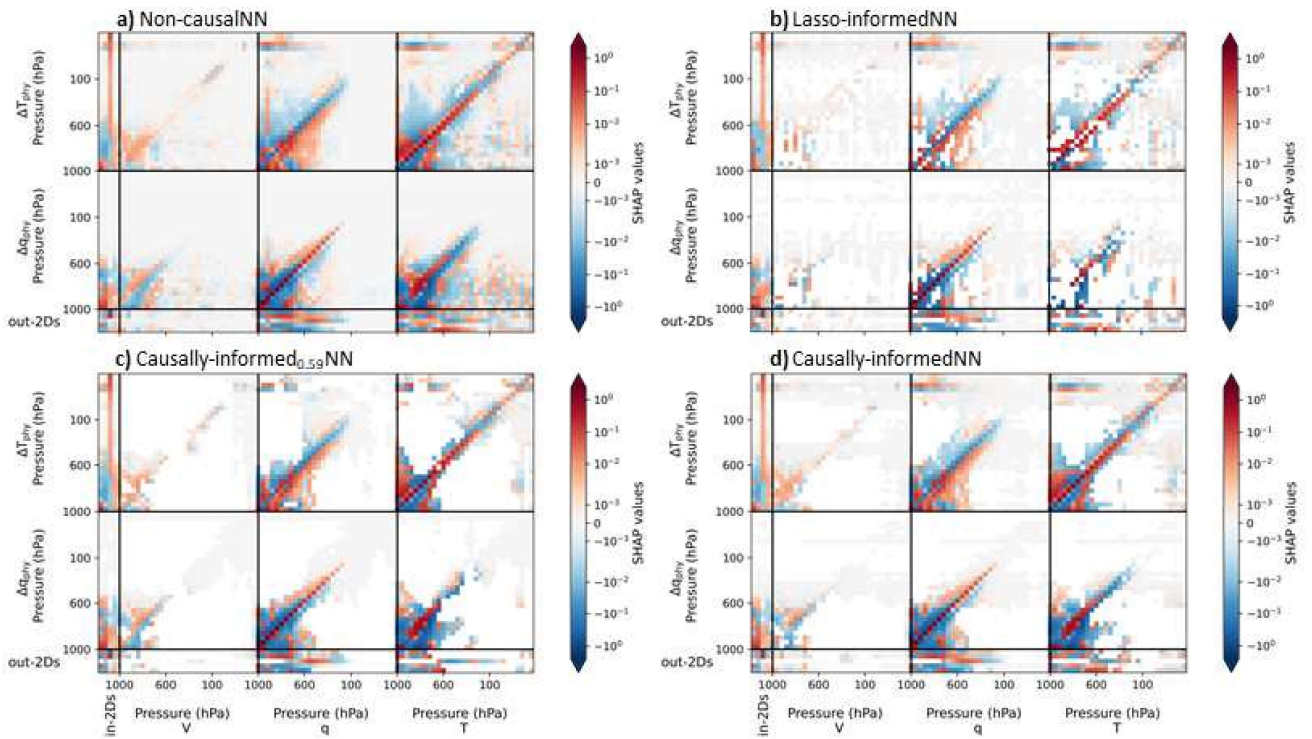

**Figure S9.** Same as Fig. 6, but including mean SHAP value sign for: **a**) Non-causalNN; **b**) Lasso-informedNN (Fig. S3b); **c**) Causally-informed$_{0.59}$NN; and **d**) Causally-informedNN.

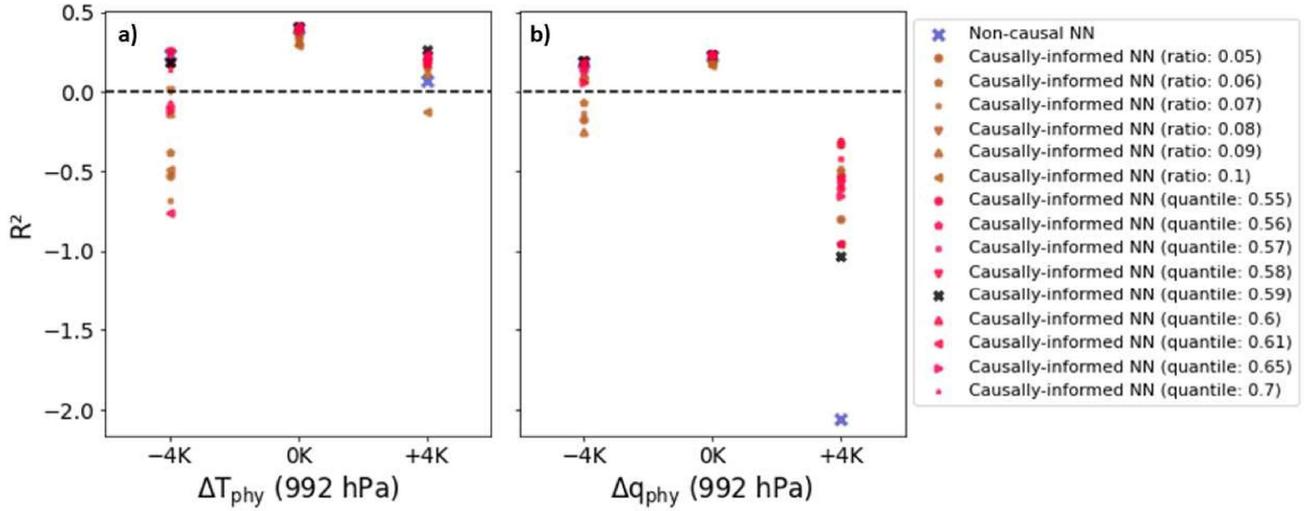

**Figure S10.** Coefficient of determination ($R^2$) of **a)** $\Delta T_{phy}$ and **b)** $\Delta q_{phy}$ at the surface (992 hPa) for the Non-causal neural network (NN) and the causally-informed NN using a number of thresholds for both approaches, ratio- (brown) and quantile-based (red). The reference SPCAM simulation (+0 K) was used for training. $R^2$ is computed using the test sets of each simulation case ($-4$ K, $+0$ K and $+4$ K). The optimal single threshold is marked with a black cross.

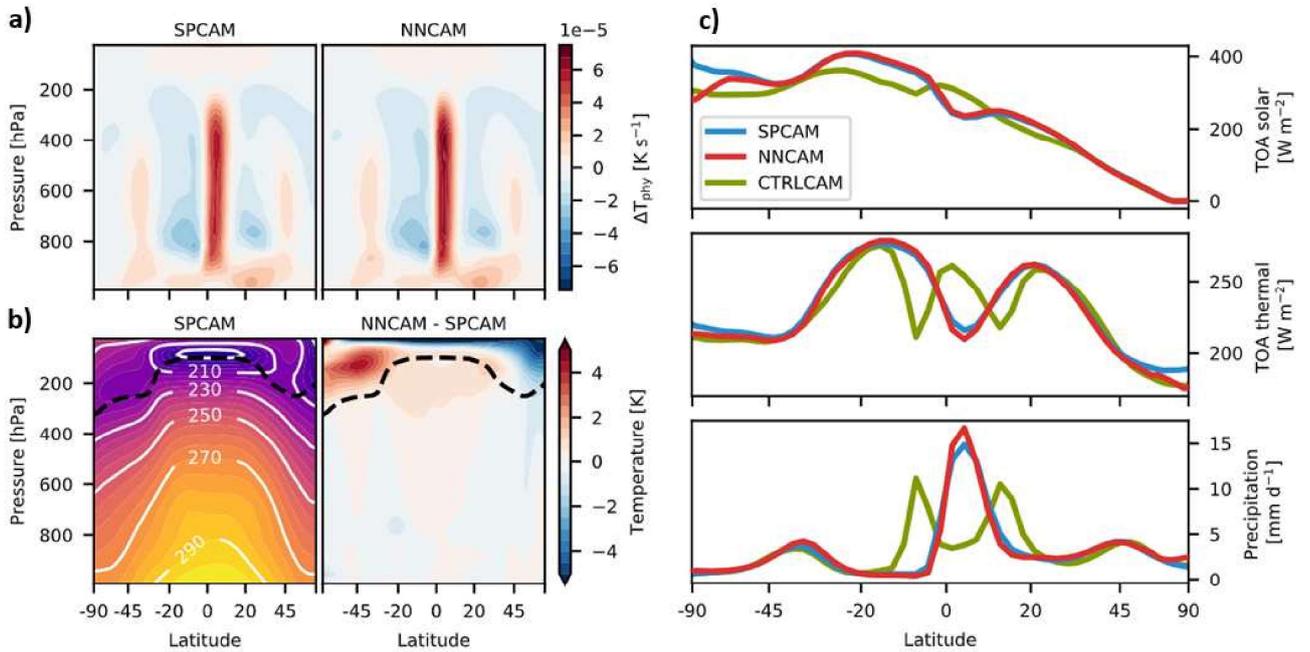

**Figure S11.** Adapted from Fig. 1 in (Rasp et al., 2018). a) Zonal mean convective and radiative subgrid heating rates $\Delta T_{phy}$. b) Zonal mean temperature T of SPCAM and NNCAM biases. Black dashed line shows the mean tropopause. c) Latitudinally resolved mean shortwave and longwave net fluxes at the top of the atmosphere and precipitation. Zonal mean values are area-weighted.